\documentclass[preprint,amsmath,amssymb,amsthm,aps]{revtex4-2}
\usepackage{graphicx}
\usepackage{capt-of}
\usepackage{lipsum}
\usepackage{dcolumn}
\usepackage{bm}
\usepackage{float}
\usepackage{xcolor}
\usepackage{placeins}
\usepackage{subeqnarray}

\usepackage{cases}

\usepackage{lineno}

\usepackage{fancyhdr}
\begin{document}

\title{Perturbation of Traveling Boussinesq Solitons by Periodic Bathymetry}
\author{A. Ludu}
\email{ludua@erau.edu}
\affiliation{Embry-Riddle Aeronautical University, Dept. Mathematics \& Wave Lab \\
Daytona Beach, FL, USA}
\author{J. Yu}
\email{jie.yu4.civ@us.navy.mil}
\affiliation{Naval Research Laboratory, Stennis Space Center, Mississippi 39529, USA}
\author{A. S. Carstea}
\email{acarst@theory.nipne.ro}
\affiliation{Department of Theoretical Physics, National Institute of Physics and Nuclear Engineering, \\
Bucharest-M\u{a}gurele 077125, Romania}
\date{\today}

\begin{abstract}
We investigate the perturbations induced by a periodic bathymetry on traveling Boussinesq solitons in a two-dimensional  configuration. We present two 
perturbation approaches to solve the  nonlinear, dispersive  and non-autonomous differential equations of the model and compare the solutions with 
numerical simulations of the original system of equations. In the approximation for small periodic corrugations we built the solutions as  modulated 
traveling waves using Fourier series. The coefficients of the series are solved  using the Green function method and the path-ordered exponential method. 
At  the second order in the relative height of bed corrugations, we obtain the perturbation as the fourth-order linear dispersive waves generated by 
the modulated traveling soliton in its wake.  In the second approach, we rewrite the Boussinesq system into a perturbed Korteweg-de Vries (KdV) 
nonlinear equation, and obtain the corresponding perturbed solitons. These analytic solutions are compared with the results of numerical simulations, 
for various parameters that characterize the effects of nonlinearity, dispersion, and bottom bathymetry. We also discuss the stability of the perturbed 
solitons in time. The perturbation approaches developed in this study are valid for any type of periodic bathymetries, and the method can readily 
be extended to non-periodic ones. 
\end{abstract}


\maketitle

\section{Introduction}

The problem of nonlinear waves and solitons propagating in water over a variable topography has long been of interest to scientists and engineers, and still remains one of the topical subjects in hydrodynamics due to its importance in coastal and oceanographic applications \cite{osborne,mei,nachh,onor,witham}. Being dispersive, surface gravity waves have finite life time, especially for localized coherent wave packets. However, large amplitude ocean waves are characterized by nonlinearity which counter-balances the dispersion and can lead to the formation of enduring localized traveling waves called solitary waves, solitons,  envelope solitons or breathers. The spontaneous formation of such stable nonlinear structures with a finite life time in the ocean, though debated in the literature, has recently gained more evidence of existence \cite{onor,china}. Ships sailing in navigation channels surrounded by shallow banks produce wakes that propagate as strongly nonlinear solitary Riemann waves for long distances \cite{1}. Such waves can result in losses of natural fine sediments and habitats along the shores,  as well as structural damages to piers \cite{3}. Freak waves, caused by nonlinear modulation or the effect of soliton fission out of tsunami waves \cite{7num,12},  are also examples of the natural formation of localized coherent structures on the ocean surface.

Nonlinear surface water waves and solitons on a flat bed are usually tractable by mathematical models based on integrable autonomous equations 
such as Korteweg-de Vries (KdV), modified KdV, or nonlinear Schr\"{o}dinger equations, which are solvable using the inverse scattering method, or other established approaches of integration \cite{ablo}. Nonlinear water wave problems in the case of variable bottom have more difficulty.   The geometrical complication brought by the non-uniform boundaries results in non-autonomous nonlinear differential equations with spatially highly variable coefficients.  
Such non-autonomous systems are often not completely integrable. When studying solitary waves in non-integrable equations, analytical techniques 
typically rely on perturbation methods, asymptotic analyses, and/or variational approximations \cite{malom}.

To address these challenges, the authors in \cite{2007Guy} proposed a new Fourier high-order spectral method to solve the full Euler equations for nonlinear waves and solitons over a large variety of topographies (e.g. periodic, rapidly varying, large amplitude), including Bragg wave-bottom resonances. This approach is based upon a Hamiltonian formulation where the Taylor series terms of the Dirichlet–Neumann operator have been rewritten to eliminate the appearance of
exponentially unbounded, hence extremely unstable operators. This generalized formulation of the problem allows even for a sudden moving bottom topography, as in the case of an earthquake-generated tsunami.

The full Euler equations governing inviscid flows can be simplified to become the reduced surface wave models that are more tractable when restricted 
to specific physical regimes, e.g., the Boussinesq-type equations which include the leading order terms describing the nonlinear and dispersive effects 
and have been shown to provide an accurate description for linear and nonlinear wave evolution in coastal regions \cite{nachh}. 
Boussinesq-type models valid for variable depth include approximations such as the mild slope hypothesis, or the terrain-following Boussinesq model \cite{nachh,nach} which can be used for a large variety of topography profiles. The problem for a two dimensional periodic seabed of arbitrary amplitude 
and shape was solved in the exact Floquet theory of linear water waves using a conformal map of the fluid region \cite{Jie2007,Jie2012,Jie2019revisit,Jie2019waveforms}. A very interesting result presented in these papers is that the wave-forms of linear time harmonic
waves on a periodic bed are frequency-dependent, and have features that are remarkably reminiscent of nonlinear wave manifestations. 
The exact Floquet theory \cite{Jie2012} provides a complete basis for water waves which can be used to solve a class of linear problems involving 
a periodic bed in confined or closed domains.

In \cite{nako} the problem of long-wave scattering by piece-wise constant periodic topography was studied for weakly nonlinear KdV solitary wave. 
An analytical approximate solution was obtained  when the length of the irregularities is larger than the characteristic nonlinear length scale. 
In that study, the wave decrement in the case of small height of the topographic irregularities is proportional to the square of 
the relative height of the topographic obstacles. An alternative form of the KdV equation with  terms depending on the bottom topography was obtained 
using the second-order perturbation approach in the weakly nonlinear, dispersive, and long wavelength limit \cite{104}. The motion of a wave with an initial 
soliton shape was studied by numerical simulations,  showing quantitative changes in the soliton’s velocity and amplitude in relation with 
bottom variations. Those authors find that the soliton’s velocity is linearly anticorrelated with the local water depth, whereas the soliton’s amplitude 
changes in a less linear manner.

In \cite{2009Craig} the authors derived appropriate Boussinesq and KdV type equations with random coefficients for weakly nonlinear dispersive waves over rapidly varying random bathymetry.  The approach consists in a perturbation theory for Hamiltonian partial differential equations, with a subsequent analysis of the random effects in the
resulting solutions. The influence of statistical properties of the bottom variations on the KdV solution’s velocity and amplitude are presented. One of the important conclusions from this study is that  coherent wave-like solutions traveling over random bathymetry persist and even maintain  their basic properties of momentum and energy transport as in the classical problem of a flat bottom.

The case of nonlinear and highly dispersive water waves on a mildly sloping bottom was analyzed using a Boussinesq type of formulation \cite{madsen0}.  
That model is based on a series expansion about  a rapidly spatially varying expansion level, in which the resulting general formulation of velocity is given as a triple-summation of terms involving high derivatives of the expansion level. In order to obtain the solutions that can be implemented in practical situations, 
those summations need to be simplified and truncated. Consequently in \cite{madsen},  a simplified  procedure  was introduced by assuming that the 
expansion level (but not the bathymetry) is slowly varying in space. For such cases  the resulting equations  include the first and second derivatives of the expansion level and up to the fifth derivatives of the velocity variables. In \cite{madsen} it was shown that the approach could be used for fully nonlinear and highly dispersive water waves interacting with a rapidly varying bathymetry.

The case of a three-dimensional generalized shallow water nonlinear equation with  time-dependent coefficients was investigated in \cite{liuos},  using 
the three-wave approach which provided exact soliton solutions.  When the bottom topography has small-scale disturbances,  it can be incorporated 
into a first order steady KdV-type of equation \cite{101}. The authors of \cite{101} used a reductive perturbation method and obtained  theoretical 
solutions representing one-soliton oscillation waves. Those solutions  become wider over a hump and narrower over a hole, indicating that the wave 
profile has the stability adjustment capability to adapt to the bottom disturbance. A different approach is used in \cite{foka},  where a new type of non-local equations (the global relations) are used to couple the appropriate transforms of all boundary values by conformal maps. Via some specific 
transformations in the Fourier space, those authors obtained an explicit Dirichlet to Neumann map, hence a final system of equations without any 
asymptotic approximations. For a recent review on the topic of nonlinear waves on a variable bathymetry,  \cite{102} can be consulted.

In the present paper we investigate the perturbations induced by a periodic bathymetry on traveling Boussinesq solitons in a two-dimensional  configuration. 
The model relies on two nonlinear, dispersive,  and non-autonomous  equations for the free surface of the water and for its surface velocity, defined in a 
conformal transformed coordinate (hereafter referred to as the mathematical plane) introduced in  \cite{Jie2019revisit,Jie2019waveforms}. 
The equations belong to an extended version of the terrain-following Boussinesq models for highly variable  bed.  These equations depend on two 
parameters describing the wave nonlinearity  and dispersion, respectively. Our goal is to study the evolution of traveling Boussinesq solitons in this 
mathematical plane,  as it is perturbed by the action of the spatially variable coefficients of the nonlinear equations, then map the resulting evolution 
back to the physical plane.  The rest of the paper is organized as follows.  Section \ref{sec2}  describes the nonlinear non-autonomous Boussinesq model.  
In Section \ref{sec3}, we develop the solutions for small corrugation-height relative to the mean water depth.  The equations for the first order perturbation  
are of Hamilton type, linear and non-homogeneous, in which  the influence of bathymetry is transferred to the non-homogeneous terms. 
The solutions are obtained using the Green function method and the path-ordered exponential method. In Section \ref{seccompare}, we show examples 
of the analytic solutions, considering a large range of values of the model parameters. We also discuss the stability of the perturbed solitons. 
We compare the approximate analytic solutions with the results of  numerical simulations of the original Boussinesq system. 
In Section \ref{sectionStefanKdV}, we describe a procedure to map the Boussinesq system under study into a perturbed KdV equation, and obtain its 
corresponding perturbed soliton. Conclusions follow in Section \ref{secconcl}. 

\section{Boussinesq model for a periodic bathymetry}
\label{sec2}

Deferring the detailed derivation to \cite{Jie2019revisit}, we here briefly describe the mathematical formulation leading to the terrain-following Boussinesq 
equations for weakly nonlinear and weakly dispersive long waves over a fully submerged, periodic seabed of arbitrary height and shape. 
The problem was formulated for a two-dimensional incompressible, irrotational and inviscid fluid layer of shallow depth. 
In the physical $(x,z)$ plane, $z=z_s(x,t)$ is the free surface and $z=-h_0 + z_b(x)$ is the seabed, where $h_0$ is the constant mean water depth, 
and $z_b(x)$ is an arbitrarily periodic function of period $\lambda_b = 2\pi/k_b$ and crest-to-trough height $\Delta z_b$. 
By definition, $z_b(x)$ has zero mean over a period, i.e., $< z_b(x) >_{\lambda_b}= 0$. The relative bed height parameter 
$\epsilon = \Delta z_b/h_0$ is not small, and for sufficiently high corrugations, $\epsilon > 1$. The zero-mean condition of the depth variation $h$ over a period of the corrugations introduces a condition for the seabed function $z_b(x)$ \cite{Jie2012,Jie2019revisit}, which establishes  a relationship between the parameters  $\epsilon, h, \beta$.

The key element of the formulation is the use of a time-independent conformal  transformation \cite{Jie2012}, which maps the 
the undisturbed flow domain $-\infty <x < \infty$, $-h_0+z_b(x) \le z \le 0$ into a uniform strip $-\infty <\xi< \infty$, $-h_m\le \eta \le 0$. The depth $h_m$ cannot be specified {\em a priori}, because the mapping is constrained by the requirement to preserve the intrinsic 
$\lambda_b$-periodicity of the problem and $h_m$ varies as the bed undulating height $\Delta z_b$ changes even if $h_0$ remains the same. 
For a fully submerged, periodic bed, $h_m < h_0$. The map is implicitly given by Fourier series in $(\xi, \eta)$ similar to the approach  used in \cite{nach,Jie2019revisit}. Since the Boussinesq system is for weakly nonlinear waves, the mathematical derivation involves Taylor-expansions of the free surface 
boundary conditions at the undisturbed surface $\eta=0$ (i.e., $z=0$), and from this point of view, a time-independent map is desirable. 
Under the map, the instantaneous free (material) surface $F(x,z,t) =z-z_s(x,t)=0$ has a conformal image $F(\xi,\eta,t) = \eta-\zeta(\xi,t)=0$. 
Note that when transforming the free surface Bernoulli equation to the mapped plane,  the potential energy remains $g z_s$; see Eq (2.4) in \cite{Jie2019revisit}. For the derivations of the terrain-following Boussinesq system details can be found in \cite{Jie2019revisit,nach}. 

The equation and exact boundary conditions for the velocity potential are  transformed into the mapped plane, then analyzed following the 
typical procedure for weakly nonlinear and weakly dispersive long waves in the sense that $\alpha \sim O(\beta^{1/2})$ and $\beta \ll 1$, where 
\begin{equation}\label{eq:alpha-beta}
\alpha = a/h_0, \quad \beta = (k_B h_0)^2,  
\end{equation}
with $a$ being a characteristic wave amplitude (e.g., half the crest-to-trough height of the surface waveform), and 
$k_B=k_b/2$ being the primary Bragg resonant wavenumber. 
Ignoring the terms of $O(\alpha\beta, \beta^2)$ and higher, a terrain-following Boussinesq system is then obtained describing the evolution of 
surface elevation $z_s$ and horizontal surface velocity $u_s\equiv u|_{z=z_s}$. These are the dimensionless equations (3.12a, b) given 
in \cite{Jie2019revisit}, and quoted here, i.e.,  
\begin{linenomath}
\begin{subequations}\label{eqt}
\begin{align}
&M\tilde{z}_{s,\tau}+[ ( h+\alpha M^{-1}\tilde{z}_s ) M\tilde{u}_s ]_{\tilde{\xi}} 
-\frac{1}{3}\beta h^2 ( M\tilde{z}_{s,\tau} )_{\tilde{\xi}\tilde{\xi}} = 0, \label{eqp1} \\
&M\tilde{u}_{s,\tau} + \tilde{z}_{s, \tilde{\xi}} +\frac{1}{2}\alpha ( \tilde{u}^{2}_s )_{\tilde{\xi}}  =0, 	\label{eqp2}
\end{align}
\end{subequations}
\end{linenomath}
where $h=h_m/h_0$ is the dimensionless depth in the mathematical plane, and  the dimensionless variables are defined as 
\begin{equation}\label{eq:normalization}
\tilde{\xi} = k_B\xi, \ \tau = \omega_B t, \ \tilde{z}_s=z_s/a, \ \tilde{u}_s = u_s/(\alpha \sqrt{gh_0}), 
\end{equation}
with $\omega_B/k_B = \sqrt{gh_0}$ and $g$ being the gravitational acceleration.  We recall that because of the zero-mean condition of the depth variation over a period of the corrugations, the $h$ parameter depends on the relative height of corrugation and the dispersion coefficient, i.e. $h=h(\epsilon, \beta)$.

The function $M(\tilde{\xi})$ is related to the Jacobian of 
transformation evaluated at the undisturbed free surface. The subscripts $\tilde{\xi}$, $\tau$ denote the partial derivatives. 
Note that $h<1$ for fully submerged periodic bed. Eqs. (\ref{eqt}) represent a non-autonomous generalization of the Boussinesq 
two-equation model for $(z_s, u_s)$ in the free-surface velocity base \cite{wu}. 
In the limiting case where the relative corrugation height $\epsilon =0$, we have $h=1$, $M=1$, and the system in Eqs. (\ref{eqt}) 
reduces to the familiar Boussinesq model for a flat bed.  Solutions to the linearized Eqs. (\ref{eqt}) are given in \cite{Jie2019revisit}, 
showing the non-sinusoidal surface waveforms for simple time-harmonic motions that reminisce nonlinear wave manifestations, 
consistent with the exact Floquet solutions for linear waves over an arbitrarily periodic bed in a finite depth \cite{Jie2007,Jie2012,Jie2019waveforms}.  

The analysis in this study is based on  Eqs. (\ref{eqt}). Hereafter the tildes for dimensionless variables are dropped for simplicity. All variables are dimensionless unless otherwise noted. From  Eq. (\ref{eqt}a), 
\begin{equation}\label{eq:Mzst}
M z_{s, \tau}= -h (M u_s)_{\xi} + O(\alpha, \beta).	
\end{equation}
Substituting back into the $\beta$-term in Eq. (\ref{eqt}a) and copying Eq. (\ref{eqt}b), we obtain a new system, i.e.,  
\begin{linenomath}
\begin{subequations}\label{eq5new}
\begin{align}
&M z_{s,\tau}+[ ( h+\alpha M^{-1}z_s ) M u_s ]_{\xi} +\frac{1}{3}\beta h^3( M u_s )_{\xi\xi\xi} = 0, \\
&M u_{s,\tau} + z_{s, \xi} +\frac{1}{2}\alpha ( u^{2}_s)_{\xi}  =0. 
\end{align}
\end{subequations}
\end{linenomath}
This has the same accuracy as the system in Eqs. (\ref{eqt}), but represents the (1+1) conservative version 
of the Boussinesq system for the evolution of surface variables of a  long wave \cite{2003zhang}.  This particular form of equations becomes useful in the following because in the limit $\epsilon =0$ (i.e. flat bottom case), the system in Eqs. (\ref{eq5new}) reduces to the Broer-Kaup (BK) nonlinear system which is integrable and has three independent Hamiltonian structures  \cite{1975Kaup,2003zhang}. Integrability for the BK system has been  proven using the inverse scattering transformation (IST), but it can also be proven using the B\"{a}cklund transformation  \cite{hirota,kaptsov}, or the Darboux transformation \cite{2000Li}. The flat-bottom Boussinesq system is also a member of the Ablowitz-Kaup-Newell-Segur hierarchy (AKNS) \cite{1975Kaup}, having exact rational
solutions relevant to the occurrence of rogue waves \cite{clarkson}, a multi-soliton solution that is expressed in a closed implicit form  \cite{2003zhang,wu,denys2007,gobbi2000}, as well as exact solutions obtained by the Painlev\'{e} method \cite{chen}. 

\section{Perturbation of a Boussinesq soliton over low corrugations}
\label{sec3}

In this section we introduce a set of solutions for the  system in Eqs. (\ref{eq5new}) in the form of  perturbations of Boussinesq solitons.  In \cite{quint}, the authors proved that the Cauchy initial problem for the non-autonomous system is well-posed and has a unique global solution following from the Gronwall fixed point inequality.  Non-autonomous  generalizations of the Boussinesq and KdV type equations with random coefficients were also studied for weakly nonlinear dispersive waves over rapidly varying random bathymetry \cite{2009Craig}.  Since Boussinesq solitons can travel in the positive and negative $\xi$ direction, we can use them to model nonlinear waves over a variable 
bottom accounting for local scattering in both directions, and locally managed dispersion.  The existence of two-way propagation of long water waves has been  studied in \cite{quint}. We consider bed corrugations of small height compared to the mean water depth $h_0$, i.e. $\epsilon<1$. Since the  bed profile is 
incorporated into Eqs. (\ref{eq5new}) through the coefficient  $M(\xi; \epsilon, \beta)$ and $h(\epsilon,\beta)$, we obtain the perturbation solutions in the expression with respect to $\epsilon$. In the limiting case of  a flat bed $\epsilon=0$ we have $M=1,h=1$  and Eqs. (\ref{eq5new}) reduce to  the traditional integrable nonlinear Boussinesq system based on the free-surface velocity, with exact solutions  $u_{sol}, z_{sol}$ \cite{1975Kaup,2000Li,2003zhang}. In  Section \ref{sec8} Appendix we give  an example of a one-soliton Boussinesq solution on a flat bed. For small amplitude bed corrugations,  we look for solutions to Eqs. (\ref{eq5new}) in the  form  
\begin{equation}\label{eq.solsol}
z_{s}=z_{sol}+\epsilon z_N, \ u_{s}=u_{sol}+\epsilon u_N, 
\end{equation}
where  $(z_N, u_N)$ are  the $\mathcal{O}(\epsilon)$ perturbations to the nonlinear soliton due to the bed, and $(z_{sol}, u_{sol})$ are the  exact soliton solution for a flat bed $(\epsilon =0)$.  

In the following we present the procedure to obtain the perturbation terms at $\mathcal{O}(\epsilon)$. Let's define 
\begin{equation}\label{eqM}  
M(\xi)=1 -\epsilon M_N(\xi). 
\end{equation}
Substituting Eqs. (\ref{eq.solsol}, \ref{eqM})  into  Eqs. (\ref{eq5new}),  we obtain a system of two linearized non-homogeneous and non-autonomous 
equations for  $u_N$ and $z_N$. Note that the $\mathcal{O}(\epsilon^0)$ terms  automatically balance since  $z_{sol}$ and $u_{sol}$ are the corresponding soliton solution in absence of corrugations. At  $\mathcal{O}(\epsilon)$ we obtain  an exact Hamiltonian system for 
\begin{equation}
z_{N,\tau}=\biggl( \frac{\delta H}{\delta u_N}\biggr)_{\xi}, \  u_{N,\tau}=\biggl( \frac{\delta H}{\delta z_N}\biggr)_{\xi}, 
\nonumber
\end{equation}
where 
\begin{equation}
H[z,u]=\int \biggl( -\frac{1}{2}z^2 -\frac{1}{2}u^2 -\alpha u_{sol} u z -\frac{\alpha }{2}z_{sol}u^2+ \frac{\beta}{6}u_{\xi}^{2} +K u+F z \biggr) d\xi. 
\nonumber
\end{equation}
The Hamiltonian equations are represented by the conservation laws
\begin{equation}\label{eq30old}
\begin{cases}
z_{N,\tau}- [K-u_N-\frac{\beta}{3}u_{N,\xi \xi}-\alpha ( z_N u_{sol}+u_N z_{sol})]_{\xi}=0, \\
u_{N,\tau}-[F-z_N-\alpha (u_N u_{sol})]_{\xi}=0,
\end{cases}
\end{equation}
where
\begin{equation}\label{sources}
\mathit{K}_{\xi}=-z_{sol, \tau} M_N-
\biggl[ u_{sol} M_{N}  +\frac{\beta}{3} (M_N u_{sol})_{\xi \xi} +h (\beta u_{sol, \xi \xi}+ u_{sol})\biggr]_{\xi}, \ \ F_{\xi}=-u_{sol,\tau} M_N.
\end{equation}
The source terms $K,F$ depend on the zeroth order soliton solutions $u_{sol}, z_{sol}$  (e.g. \cite{2000Li}) and the corrugation geometry.  Eqs. (\ref{eq30old}) are linear dispersive non-homogeneous and non-autonomous wave equations, in which the functionals $\mathit{K}_{\xi}, \mathit{F}_{\xi}$ play the role of distributed wave sources, or virtual current densities in the terminology of Hamiltonian dynamics. These non-homogeneous terms combine the wave nonlinearity with the effect of bathymetry via the product between the flat-bed soliton solutions $(z_{sol}, u_{sol})$ and the Jacobian $M_N(\xi)$ \cite{Jie2019revisit}. These Hamiltonian conservation laws are in agreement with the perturbation theory approach for Hamiltonian partial differential equations in \cite{2009Craig}  for the analysis of the bathymetry effects in the resulting solutions.

\subsection{The $\mathcal{O}(\epsilon)$ perturbation solutions}
\label{subsec1}
	
To obtain $z_N, u_N$ we solve  Eqs. (\ref{eq30old}) choosing a family of Boussinesq one-soliton traveling solutions $ z_{sol}(\xi-v\tau), u_{sol}(\xi-v\tau)$ \cite{2003zhang} given by Eqs. (\ref{soliton}) in Appendix. These solitons depend on the two parameters $\alpha, \beta$ of the problem,  and  the normalized soliton velocity $v=v_{sol}=1+\alpha^2/2$ which depends only on the normalized soliton amplitude.  We choose a certain traveling interval $I$ along the $\xi-$axis, extending over many times soliton width $L_{sol}$. Homogeneous boundary conditions are applied for $(z_N, u_N)$ and for their first order $\xi-$derivatives  at the two ends of this $I$ interval. We assume that the initial unperturbed soliton begins to travel  over quiescent water, thus choosing the  homogeneous initial conditions for Eqs. (\ref{eq30old}). We study the $\mathcal{O}(\epsilon)$ solutions $z_N,u_N$  only during the time interval when the soliton travels inside the interval $I$. The solutions obtained at this order are influenced by the bathymetry via the source terms $K,F$. These solutions are non-zero only in the region where the soliton $(z_{sol},u_{sol})$ has  non-zero contribution, i.e. only when $\xi$ belongs to the support of the soliton profile. Indeed, away from this region the soliton profile drops rapidly to zero and thus the source  terms in Eq. (\ref{sources}) vanish. Without the source terms,  and since the  boundary  and initial condition are both homogeneous, the  solution $(z_N,u_N)$ approaches zero outside the soliton support. 
In other words, the $\mathcal{O}(\epsilon)$ perturbation  travels with the  soliton,  modulating its envelope.

The traditional approach to solve Eqs. (\ref{eq30old}) is to  build the solution in terms of Fourier series in  $\xi,\tau$.  Because the  $\mathcal{O}(\epsilon)$ perturbation solution  $(z_N,u_N)$  emerges on top of a traveling soliton, we introduce the co-moving  coordinate $\bar{\xi}=\xi- v \tau-\xi_0$ and the non-singular change of variables  $(\xi,\tau)\rightarrow (\xi,\bar{\xi})$.  
For Lipschitz continuous solutions \cite{lips} with evolution restricted to the bounded region $I$ around the traveling soliton $(z_{sol},u_{sol})$, such a change of variables in the  Fourier series is just a coordinate re-scaling and does not alter the convergence properties of the series \cite{convergFS,nach}.   The parameter $\xi_0$ is chosen such that the  initial condition $\bar{\xi}=0$ takes place at the front of the traveling soliton, to secure the retarded causal solutions for $(z_{N},u_{N})$. Thus, at $\mathcal{O}(\epsilon)$ the solution can be represented by the series 
\begin{equation}\label{eb1}
z_N(\xi,\bar{\xi})=\sum_{n=-\infty}^{\infty}z_{n}(\bar{\xi})e^{i n  \xi}, \ u_N(\xi,\bar{\xi})=\sum_{n=-\infty}^{\infty}u_{n}(\bar{\xi})e^{i n  \xi}. 
\end{equation}
For  the bathymetry contribution in Eqs. (\ref{sources}) we use the Fourier expansion
\begin{equation}\label{eb5m}
M_N (\xi)=\sum_{n=-\infty}^{\infty} M_n e^{i n  \xi},
\end{equation}
where $M_n$ are the Fourier coefficients  $2 \pi M_n=\int_{-\pi}^{\pi} M_{N}(\xi) \exp (-i n \xi) d\xi$.
Substituting Eqs. (\ref{eb1},\ref{eb5m}) into Eqs. (\ref{eq30old},\ref{sources}), we  proceed to solve  the resulting system for each $n$, imposing 
zero boundary conditions for $z_N$ and $u_N$, as well as  their first order $\xi-$derivatives, at the ends of the interval $I$. 

\subsubsection{Equations for $n=0$}
\label{subsec2}

This part of the solution depends only on  $\bar{\xi}$. It  does not explicitly includes  the corrugation term $M_N$. 
From  Eqs. (\ref{eq30old}) we have
\begin{equation}\label{eq30oldnzero}
\begin{cases}
-v z_{0}^{'}+u_{0}^{'}+\frac{\beta}{3}u_{0}^{'''}+\alpha ( z_{0} u_{sol}+u_0 z_{sol})^{'}=-h (\beta u_{sol}^{'''}+ u_{sol}^{'}), \\
-v u_{0}^{'}+z_{0}^{'}+\alpha (u_{0} u_{sol})^{'}=0,  
\end{cases}
\end{equation}
where primes represent the differentiation in $\bar{\xi}$.
Integrating once  and noting that  $z_0=(v -\alpha  u_{sol})u_{0}$, we obtain from the first  equation in Eqs. (\ref{eq30oldnzero}) 
\begin{equation}\label{25}
\frac{\beta}{3}u_{0}^{''}+(1+\alpha z_{sol}+2 \alpha v u_{sol}-\alpha^2 u_{sol}^{2}-v^2) u_{0}=-h (u_{sol}+\beta u_{sol}^{''}).
\end{equation}
Here the integration constants have been canceled, since  we consider only the localized solutions which approach zero at the 
boundaries of $I$. From  Eq. (5b), applying for the flat bed soliton solution and integrating once, we have 
$-v u_{sol}+z_{sol}+\alpha u_{sol}^{2}/2=0$. This reduces Eq. (\ref{25}) to
\begin{equation}\label{26}
\frac{\beta}{3}u_{0}^{''}+(1-v^2 + 3 \alpha z_{sol}) u_{0}=-h (u_{sol}+\beta u_{sol}^{''}).
\end{equation}
The general solution to Eq. (\ref{26}) is given by a sum of the general solution of the homogeneous part of the equation (which we denote as $u_0^{hom}$) 
and a particular solution of the fully non-homogeneous equation. The latter can be obtained using the Green function method \cite{1912,1927}. The homogeneous solution can be obtained by converting Eq. (\ref{26}) into a standard Volterra integral equation with a  continuous kernel \cite{volt}
\begin{equation}
u_0^{hom}(\bar{\xi})+3 \alpha \int_{0}^{\bar{\xi}} (\bar{\xi}-s) z_{sol} (s) u_0^{hom}(s)ds=0.
\nonumber
\end{equation}
For this homogeneous Volterra integral equation there are  many approaches to obtain its solution, including the Taylor's expansion method.  However, because the variable coefficient $1-v^2 + 3 \alpha z_{sol}=\alpha (-\alpha/2 +3 z_{sol})$ in Eq. (\ref{26}) becomes negative when $z_{sol}<\alpha/6$,  some homogeneous solutions  $u_0^{hom}$ may become unstable in the sense of Ulam \cite{ulam}.  To obtain the non-homogeneous part of the solution $u_0$ we employ the expressions for $z_{sol}, u_{sol}$ and $v$ in Eq. (\ref{soliton}) and use Mathematica\textsuperscript{\textregistered} symbolic exact integration to obtain analytic forms for the 
Green function  $G(\bar{\xi}, \bar{\xi}^{\ '})$ for Eq. (\ref{26}), i.e.,  
\begin{equation}\label{26p}
G=(1+e^{i X})^4 e^{-3 \alpha \sqrt{43} |X|} 
\begin{cases}
_{2}F_{1}(4,4 + 6 \sqrt{4+3\alpha^2},1+ 6\sqrt{4+3 \alpha^2}, -e^{-X}), \ \hbox{if} \   X \ge 0  \\
_{2}F_{1}(4,4 - 6 \sqrt{4+3\alpha^2},1- 6\sqrt{4+3 \alpha^2}, -e^{X}), \ \hbox{if}  \  X \le 0,
\end{cases}
\end{equation} 
where $X(\bar{\xi},\bar{\xi}') =\alpha \sqrt{3(4+\alpha^2)/4 \beta}(\bar{\xi}-\bar{\xi}^{\ '})$ and $_{2}F_{1}$ is the confluent hypergeometric function.  This Green function is continuous on any symmetric and bounded interval $I$, and for any $\alpha>0$ (see for example Eq. (7) in \cite{1912}). Furthermore, its second derivative is the delta Dirac distribution.   We keep  the  $\mathcal{O}(\alpha^2)$ terms formally in Eqs. (\ref{25},\ref{26})  because the Green function is not defined at $\alpha=0$, yet it is integrable. The solution to  Eq. (\ref{25}) is then given by the Green's second formula  \cite{1912,1927}, i.e.,
\begin{equation}\label{27}
u_{0}(\bar{\xi})=u_0^{hom}(\bar{\xi})-h \int_{I} G(\bar{\xi},\bar{\xi}^{\ '}) [u_{sol}(\bar{\xi}^{\ '})+\beta u_{sol}^{''}(\bar{\xi}^{\ '})]d \bar{\xi}^{\ '}. 
\end{equation}
It then follows 
\begin{equation}\label{27b}
z_{0}(\bar{\xi})=(v -\alpha  u_{sol}-1) \biggl( u_0^{hom}-h\int_{I} G(\bar{\xi},\bar{\xi}^{\ '}) [u_{sol}(\bar{\xi}^{\ '})+\beta u_{sol}^{''}(\bar{\xi}^{\ '})]d \bar{\xi}^{\ '}\biggr). 
\end{equation}
For $\epsilon < 1$, $h$  in Eqs. (\ref{27}-\ref{27b}) can be approximated as $h=1+\mathcal{O}(\epsilon^2)$; see the example in Section \ref{seccompare}. This means that the influence of small corrugations does not appear in the $n=0$ component of the solution at $\mathcal{O}(\epsilon)$. Nevertheless, analytical expressions for the homogeneous solution $u_0^{hom}$ can be obtained up to order $\mathcal{O}(\alpha)$, and the integrals in Eqs. (\ref{27},\ref{27b})  can be evaluated exactly, resulting in combinations of hyperbolic, inverse hyperbolic, and polylogarithm functions. These expressions are too lengthy to present here, but the details of the calculations and the exact forms of the solutions in Eqs. (\ref{26}-\ref{27b}) for various parameter values are provided in the Mathematica\textsuperscript{\textregistered} file "Supplemental Material A."

\subsubsection{Equations for $n \neq 0$}
\label{subsec3}

Collecting the terms associated with $e^{in\xi}$, $n\neq 0$, we obtain the equations for $z_n(\bar{\xi})$ and $u_n(\bar{\xi})$ from  Eqs. (\ref{eb1},\ref{eb5m}). These are  the third order differential equations. Let the  vector $Y_n (\bar{\xi})=[u_n,z_n,q_n,p_n]^{T}$, where $T$ stands for the transpose, and  
$q_n (\bar{\xi})=d u_{n}/d\bar{\xi}$, $p_n(\bar{\xi})=dq_{n}/d\bar{\xi}$. We transform the third order differential equations into  a non-homogeneous, non-autonomous system of first order differential equations 
\begin{equation}\label{eb2}
 Y^{'}_{n}=\mathcal{A}_n Y_n+\mathcal{B}_n
\end{equation}
with the associated   Cauchy initial condition $Y_n(0)=Y_{n0}$. The matrices are defined as follows.
\begin{equation}\label{eb3}
\mathcal{A}_n(\bar{\xi})=
\begin{pmatrix}
0 & 0 & 1 & 0 \\
-\alpha (u_{sol}^{'}+i n \beta u_{sol}) & -i n \beta  & v-\alpha u_{sol} & 0 \\
0 & 0 & 0 & 1 \\
{A}_{n}^{41} & 
-3 i n v-\frac{3\alpha}{\beta} u_{sol}^{'} & {A}_{n}^{43} & -3 i n \beta 
\end{pmatrix}
\end{equation}
where 
\begin{equation}
{A}_{n}^{41}=-3in-3 i n \alpha(v  u_{sol}+ z_{sol} )-\frac{3\alpha}{\beta}(v u_{sol}^{'}+  z_{sol}^{'}-\alpha u_{sol}u_{sol}^{'})+3 i n \alpha^2 u_{sol}^{2}+i \beta^3 n^3,
\nonumber
\end{equation}
\begin{equation}
{A}_{n}^{43}=\frac{3(v^2-1)}{\beta} -\frac{3\alpha}{\beta}( 2 v u_{sol}+ z_{sol}-\alpha u_{sol}^{2})+3n^2 \beta^2.
\nonumber
\end{equation}
and $\mathcal{B}_n(\bar{\xi})=M_n [0,-v u_{sol}^{'},0,B_n^4]^{T}$, where 
\begin{equation}\label{eb4}
B_n^4= (i n^3 \beta^3 -3 i n) u_{sol}+3\biggl( n^2 \beta^2 -\frac{1-v^2+\alpha v u_{sol}}{\beta } \biggr) u_{sol}^{'}-3 i n \beta u_{sol}^{''}-u_{sol}^{'''}+\frac{3v}{\beta}z_{sol}^{'}.
\end{equation}
The unique solution to the non-autonomous system in Eqs. (\ref{eb2}-\ref{eb4}),  for each $n$, is the path-ordered exponential (PE) \cite{2014Giscard}
\begin{eqnarray}\label{eqseries}
&&
Y(\bar{\xi})=Y_0+\int_{0}^{\bar{\xi}} \mathcal{A}(s_1) ds_1 Y_0 +\int_{0}^{\bar{\xi}} \mathcal{A}(s_1) \biggl( \int_{0}^{s_1}  \mathcal{A}(s_2) ds_2 \biggr) ds_1  Y_0
\nonumber\\
&& + \sum_{k=3}^{\infty} \biggl[ \int_{0}^{\bar{\xi}}  \int_{0}^{s_1}\cdots \int_{0}^{s_{k-1}} \mathcal{A}(s_1) \mathcal{A}(s_2) \cdots \mathcal{A}(s_k) ds_{k} ds_{k-1}\cdots ds_1 \biggr] Y_{0} \nonumber\\
&&\qquad  +\sum_{k=1}^{\infty} \int_{0}^{\bar{\xi}}  \int_{0}^{s_1}\cdots \int_{0}^{s_{k-1}} \mathcal{A}(s_1) \mathcal{A}(s_2) \cdots \mathcal{A}(s_{k-1}) \mathcal{B}(s_k) ds_{k} ds_{k-1}\cdots ds_1,
\label{eb5}
\end{eqnarray}
where the subscript $n$ is dropped for the sake of neatness. The products $\mathcal{A} \cdots\mathcal{A} \mathcal{B}$ are understood as 
the matrix products, and the dots represent sequences of iterated integrals from $1$ to $k$. With the homogeneous initial condition $Y(\bar{\xi}=0)=Y_0=0$  Eq. (\ref{eb5}) becomes 
\begin{equation}\label{eb5p}
Y(\bar{\xi})= \int_{0}^{\bar{\xi}} \biggl[ \mathcal{B}(s_1)  +  \int_{0}^{s_1} \mathcal{A}(s_1)\mathcal{B}(s_2)ds_2 +  \int_{0}^{s_1}\int_{0}^{s_2} \mathcal{A}(s_1)\mathcal{A}(s_2)\mathcal{B}(s_3)ds_3 ds_2+\dots \biggr] ds_1.
\end{equation}
Since the goal of the calculations here  is to obtain the  perturbation to the soliton profile and velocity, we only need components  $Y_{n,2}=z_n$ and  $Y_{n,1}=u_n$. From  Eq. (\ref{eb5p}) 
\begin{eqnarray}
	&& z_n(\bar{\xi}) = \int_{0}^{\bar{\xi}} \mathcal{B}_{n,2}(s_1)  ds_1 +   \int_{0}^{\bar{\xi}} ds_1 \int_{0}^{s_1} \mathcal{A}_{n, 22}(s_1)\mathcal{B}_{n,2} (s_2)ds_2 \nonumber\\
	&& \qquad + \int_{0}^{\bar{\xi}}  ds_1 \int_{0}^{s_1} ds_2 \int_{0}^{s_2} [\mathcal{A}_{n,22}(s_1)\mathcal{A}_{n,22}(s_2)\mathcal{B}_{n,2}(s_3)+\mathcal{A}_{n,23}(s_1) \mathcal{B}_{n,4}(s_3)]ds_3 +\dots 
	\label{eb5pp}
\end{eqnarray}
\begin{eqnarray}
	&& u_n(\bar{\xi}) = \int_{0}^{\bar{\xi}}  ds_1 \int_{0}^{s_1} ds_2 \int_{0}^{s_2} \mathcal{B}_{n,4} (s_3)ds_3  \nonumber \\
	&& \qquad + \int_{0}^{\bar{\xi}} ds_1 \int_{0}^{s_1} ds_2 \int_{0}^{s_2} ds_3 \int_{0}^{s_3}  [\mathcal{A}_{n,42}(s_3)\mathcal{B}_{n,2}(s_4)+\mathcal{A}_{n,44}(s_3) \mathcal{B}_{n,4}(s_4)]ds_4 +\dots 
	\label{eb5ppp}
\end{eqnarray}
The convergence of the PE iterated series can be handled through various methods of decomposition \cite{1998Lam,2014Giscard,2019Giscard}. One criterion of convergence of the PE series is that  the matrix $\mathcal{A}$ is sparse \cite{OEpseudo}, and another criterion of convergence is to ask the integrand functions to be square integrable with norms less than $\pi$ (see Eq. (67) in \cite{2009Blanes}). For a recent review of the theory see also \cite{2009Blanes}. The convergence of the  series in Eqs. (\ref{eb5}-\ref{eb5ppp}) is guaranteed by both these criteria. First,  each of the real entries in matrices $\mathcal{A}, \mathcal{B}$ contains, as factors, the smallness parameters $\alpha, \beta, v^2-1=\mathcal{O}(\alpha^2)$, with two exceptions: the element $\mathcal{A}_{2,3}=v-\alpha u_{sol}$ and the last term of the last component of $\mathcal{B}_n$. Nevertheless,  because of the positions of the zero entries in the $\mathcal{A}, \mathcal{B}$ matrices, to retain the  terms up to orders $\mathcal{O}(\alpha)$ and $\mathcal{O}(\beta)$ in $Y_{n,1}$ and $Y_{n,2}$,   it is sufficient to keep only the first three terms in Eqs. (\ref{eb5pp},\ref{eb5ppp}), as all subsequent higher-iteration terms have smaller orders. Secondly,  it is known that even if $\mathcal{A}_n$ are completely dense (which is the case here since  a $4 \times 4$ matrix cannot be sparse) but if it has rapidly decreasing entries away from the diagonal, the corresponding PE series still converges. Matrices of this type are called pseudosparse \cite{OEpseudo}. This is  exactly our case because all real entries are functions of either the localized soliton $(z_{sol},u_{sol})$ or their derivatives. That is,  the matrices $\mathcal{A}_n$ in Eq. (\ref{eb3}) are pseudosparse. In the working approximation, Eqs. (\ref{eb5pp},\ref{eb5ppp}) reduce to the following perturbations to the soliton profile
\begin{eqnarray}
&& z_n (\bar{\xi})=3 \beta\int_{0}^{\bar{\xi}}ds_1 \int_{0}^{s_1} z_{sol}(s_2) ds_2+\mathcal{O}(\alpha^2,\beta^3),
\label{x200}
\end{eqnarray}
\begin{eqnarray}
&& u_n (\bar{\xi})= \beta u_{sol}(\bar{\xi})-3 \beta^2 \int_{0}^{\bar{\xi}} (1-u_{sol}(s_1))ds_1 \int_{0}^{s_1} z_{sol}(s_2) ds_2 +\mathcal{O}(\alpha^2,\beta^3). 
\label{x201}
\end{eqnarray}
We also mention that all iterated terms in the path-ordered exponential (PE) in Eqs. (\ref{eb5pp},\ref{eb5ppp}) can be integrated analytically exact, and they result in finite sums of hyperbolic tangent, or polylogarithm Li$_{k}(\bar{\xi})$ functions of order $k$ multiplied by powers of the soliton half-width $L_{sol}^k$. All these expressions are bounded from above by convergent sequences $L_{sol}^k/(2k)!$ , a fact that guarantees the convergence of the PE iterated integrals.

While the calculations described above require careful attention,  the advantage  is that it represents a workable analytic solution,  valid at  $\mathcal{O}(\epsilon)$, for the evolution of a Boussinesq soliton perturbed by the small bathymetry. The analytical calculations for the integral solutions Eqs. (\ref{x200},\ref{x201}), and the sequence of the maximal values of these integrals (for convergence purposes)  are provided together with other examples, in the Mathematica\textsuperscript{\textregistered} file "Supplemental Material A."

The laboriousness of the procedure to obtain the solutions $(z_N, u_N)$, even at the first order in $ \epsilon$, is justified by the fact that non-autonomous differential equations such as Eqs. (\ref{eq30old},\ref{eq30oldnzero}) give rise to processes depending on two factors: the initial moment of time and initial conditions. This is in
contrast to autonomous differential equations which have solutions  depend only on the initial conditions. For an autonomous nonlinear Boussinesq system on a  flat bottom, we can define a global attractor which is a subset of the phase space of solutions that captures the asymptotic behavior of the system. In the case of non-autonomous systems (e.g. the nonlinear Boussinesq system on a variable  bathymetry), we need to calculate (or prove the existence/non-existence) of a new, specific pullback attractor. Since  the contribution of bathymetry to the traveling nonlinear waves can be interpreted  as a time-dependent external interaction, one could study a pullback attractor for this system by   identifying the time-dependent invariant sets that attract all
solutions initialized in the remote past. Moreover, since the existence of a global pullback attractor for weakly dispersive systems, such as ours, was rigorously demonstrated in the literature \cite{pba}, one could follow this path to determine more quantitative features of the coupling between bathymetry and nonlinear waves. But this is beyond the scope of this paper.

\subsection{Asymptotic scattered solutions}
\label{subsec4}

Outside the region of the localized support of soliton, $z_{sol}\simeq 0$,  $u_{sol} \simeq 0$.  Thus,  the equations for $z_N$ and $u_N$ are decoupled, 
and Eqs. (\ref{eq30old}) reduce to the linear dispersive wave equations
\begin{equation}\label{eqL1}
z_{\tau \tau}-z_{\xi \xi}-\frac{\beta}{3}z_{\xi \xi \xi \xi}=0, \ \ u_{\tau \tau}-u_{\xi \xi}-\frac{\beta}{3}u_{\xi \xi \xi \xi}=0,	
\end{equation}
and $u_{\tau}=-z_{\xi}$, together with the dispersion relation
\begin{equation}\label{eqdisp}
\omega(k)=\pm k\sqrt{1- \frac{\beta k^2}{3}  }.
\end{equation}
The wave velocities in the mathematical plane are  
\begin{equation}\label{eqvv}
v_{ph}=\sqrt{1- \frac{\beta k^2 }{3} }<1, \ \ v_{gr}=\frac{3-2 \beta k^2}{\sqrt{9- 3 \beta k^2 }}<1.
\end{equation}
We notice that both velocities are less than the group velocity $v_{sol}$ of a single soliton on a flat bed as given in  Eq. (\ref{soliton}), i.e. $v_{ph} <1< v_{sol}=1+\alpha^2/2$. This  means that the scattered waves generated in the wake of a traveling soliton  do not tend to overtake the soliton leading front, unless the corrugations are so high that  the perturbed soliton moves slower in  an effective layer of shallower depth. Adjacent to the region of soliton support, the coefficient functions in Eq. (\ref{eq30old}) and the non-homogeneous source terms in Eq. (\ref{sources})  all cancel out.  As a result, $(z_{N},u_{N})$
represent the scattered waves in the asymptotic region, which lies outside the soliton's support. In fact, the asymptotic solutions $(z_{N},u_{N})$ belong to the class of solutions of the linearized Boussinesq system, which is derived from Eqs. (\ref{eq5new}) by setting $\alpha=0$. These solutions are similar to those obtained and discussed in detail in reference \cite{Jie2019revisit,Jie2019waveforms}, and are in agreement with the formation of scattered wave fields, and the stability of the principal coherent solution obtained in \cite{2009Craig}.

To integrate the 
equations for $(z_N, u_N)$ in the asymptotic region, the  matching continuity condition is applied between these solutions and the trailing front of the perturbed soliton. 
Sommerfeld radiation boundary conditions are also applied at the left end of these linear solutions (assuming right-moving waves).  
Even if the contribution of the corrugated bed does not explicitly  occur in these equations, $(z_N,u_N)$ are still affected by the   bed corrugations 
because of the matching conditions at the  trailing front of the perturbed  soliton.

To build these solutions, we follow  \cite{Jie2019revisit,Jie2019waveforms} and represent the solutions in a wave packet using the linearized dispersion relation Eq. (\ref{eqdisp}), i.e.,
\begin{equation}
z_{N}(\xi,\tau)=\int Z(k) e^{i(k \xi-\omega(k) \tau)} dk, \ \ u_{N}(\xi,\tau)=\int U(k) e^{i(k \xi-\omega(k) \tau)} dk.
\nonumber
\end{equation}
We impose the matching condition between the right bound of this wave packet and a point of minimal amplitude $z_{min}$ chosen on the trailing front of the traveling soliton, i.e.,  $z_{min}(\tau)=z_{sol}(\xi=v \tau, \tau)$. Following this moving boundary condition, and leaving a free boundary condition 
at the left bound of the wave packet,  we obtain the linear wave solution for the wake of the soliton
\begin{equation}\label{linaa}
z_{N}=
\begin{cases}
\displaystyle{ \frac{z_{min}}{v-1}+\frac{v z_{min}}{1-v^2}e^{i\sqrt{\frac{3(1-v^2)}{\beta}}(\xi-v \tau)}} \ \ \ \hbox{if} \ \biggl|\xi-\xi_{init}+\frac{L_{sol}}{2}\biggr|<v \tau, \\
0, \ \hbox{otherwise},
\end{cases}
\end{equation}  
where $\xi_{init}$ is the initial position of the soliton at $\tau=0$, and $L_{sol}$ is the half-width of the soliton. Implementing $u_{N,\tau}=-z_{N,\xi}$ from Eq. (\ref{eqL1}), and using the matching continuity condition  between the scattered wave velocity and the velocity of trailing front of the soliton $u_{N}(\xi=v \tau, \tau)=v_{sol}=1+\alpha^2/2$ it results
\begin{equation}\label{linaa2}
u_{N}=
\begin{cases}
\displaystyle{ \frac{2+\alpha^2}{2}+\frac{z_{min}}{v^2-1}\biggl( e^{-\frac{i(L_{sol}+2\xi_{init})}{2}\sqrt{\frac{3(1-v^2)}{\beta}}}-e^{i\sqrt{\frac{3(1-v^2)}{\beta}}(\xi-v\tau)} \biggr) }  \ \ \ \hbox{if} \ \biggl|\xi-\xi_{init}+\frac{L_{sol}}{2}\biggr|<v \tau, \\
0, \ \hbox{otherwise}.
\end{cases}
\end{equation}  
For $v<1$ we obtain the  monochromatic traveling waves, and for $v\ge 1$ we obtain the exponentially attenuated  waves. The  waves in Eq. (\ref{linaa}) have a  constant wavelength because the coefficients of the governing differential equations are $\pi-$periodic in both the mathematical and physical plane, by the construction of the conformal map. These results are in agreement with the description of trailing sine waves in \cite{chang0} (for example Figs. 7, 9 in that paper).

The solution Eq. (\ref{linaa})  has a good behavior because it represents the waves that do not overtake the soliton leading front. Moreover, the wavelength prescribed by the dispersion relation Eqs. (\ref{eqL1},\ref{eqdisp}) has a value  $\lambda \simeq \sqrt{4 \beta / 3 \alpha^2 }\pi$ which is close to  the wavelength of the linear solution Eq. (\ref{linaa}) for a wide range of values of the parameters $\alpha, \beta \in [0.2,0.6]$, especially if $\alpha \sim \beta +0.2$.  

The solution Eq. (\ref{linaa}) depends on the periodicity of $M$ and  the condition at  the right moving boundary (i.e.  the trailing front of the soliton). The solution is expressed as a wave packet in which each component  has two parameters: the amplitude and  wavelength. The phase velocity is imposed by the soliton motion. Because the equations in Eq. (\ref{eqL1}) have the fourth-order derivatives in $\xi$,  there should be two matching conditions at the connection point with the soliton. This means that  the solution is a linear combination of two wavelengths. This observation is confirmed by the results from our  numerical simulations (see Section \ref{seccompare}),  showing the occurrence of dominant wavelengths $\lambda \sim \pi$ and $2 \pi$ in the wake of the soliton.  This finding is similar to the results  in \cite{Jie2019waveforms} showing the occurrence of traveling waves with wavelengths being the integer multiples of the bathymetry period.

\section{Results and discussions}
\label{seccompare}

For an arbitrary form of periodic bed corrugations, one must construct the corresponding conformal map, e.g., following the method in 
\cite{Jie2012}, in order to obtain the coefficient function $M(\xi)$ from the Jacobian of transformation. 
For the examples of solutions to be presented in this section, we use a particular family of bed forms for which the associated 
conformal map is known, i.e.,  
\begin{subeqnarray}\label{eq:cusp-bed-map}
 x &\!\!=\!\!& \xi -\epsilon h \beta^{1/2}\frac{\cosh{(2\beta^{1/2}\eta)}}{\sinh{(2\beta^{1/2} h)}} \sin{2\xi},\slabel{eq:x-map}\\
 z &\!\!=\!\!& \eta -\epsilon h \,\frac{\sinh{(2\beta^{1/2}\eta)}}{\sinh{(2\beta^{1/2} h)}} \cos{2\xi}. \slabel{eq:z-map}
 \end{subeqnarray}
Since the map is chosen {\em a prior}, $z_b(x)$ is subsequently obtained from the map via an implicit expression. 
In dimensionless form, the bed is at $z=-1+z_b(x)$ and mapped into $\eta=-h$. Thus, from Eq. (\ref{eq:cusp-bed-map}) with  $\eta=-h$,
\begin{equation} 
z_b(x) = (1-h) + \epsilon h \cos{2\xi}, \quad \mbox{where }\  x = \xi -\beta^{1/2} \epsilon h \coth{(2\beta^{1/2} h)} \sin{2\xi}.  
\label{eq:zb}
\end{equation}
The relative height is limited, 
\begin{equation}
\epsilon \leq \frac{\tanh{(2\beta^{1/2} h)}}{2\beta^{1/2} h}.  
\label{eq:cusp-bed-epsilon} \end{equation}
Eqs (\ref{eq:cusp-bed-map})--(\ref{eq:cusp-bed-epsilon}) are given in Appendix in \cite{Jie2019revisit}.  
By requiring $z_b(x)$ to have zero average (see Section II),  we obtain from Eq. (\ref{eq:zb}) the condition to determine the depth $h$ in the 
mathematical plane, i.e., 
\begin{equation}
 h= 1- \epsilon^2 \beta^{1/2} h^2 \coth{(2\beta^{1/2} h)}.
\label{eq:cusp-bed-h}\end{equation}
The profile $z_b(x)$ in Eq. (\ref{eq:zb}) is nearly sinusoidal for small $\epsilon$, but becomes cusp-like as $\epsilon$ approaches the limit 
in Eq. (\ref{eq:cusp-bed-epsilon}). 
By definition, $M(\xi) = (J|_{\eta=0})^{1/2}$, where $J$ is the Jacobian of transformation. Since ${\partial x }/{\partial \eta} =0$ at $\eta=0$, 
we therefore have 
\begin{equation}
 M(\xi) =\left. \frac{\partial x }{\partial \xi }\right |_{\eta=0}= 1-\epsilon\frac{2\beta^{1/2} h}{\sinh{(2\beta^{1/2} h)}}\cos{2\xi}\equiv 1-\epsilon M_N. 
\label{eq:cusp-bed-M}
\end{equation}
From Eq. (\ref{eq:x-map}),  at $\eta=0$ 
\begin{equation}
 x = \xi -\epsilon h \,\frac{\beta^{1/2}}{\sinh{(2\beta^{1/2} h)}}\sin(2\xi),    
\label{eq:x-xi-at-eta0}
\end{equation} 
using which we can rewrite the solutions $z_s(\xi)$ and $u_s(\xi)$ as functions of $x$ at a given time.

For each example we choose the nonlinearity parameter $\alpha$ which describes soliton relative amplitude, the dispersion parameter $\beta$, and 
the relative height $\epsilon$ of the corrugations. Note that the values of $\beta$ and $\epsilon$ affect the conformal map Eq. (\ref{eq:cusp-bed-map}), hence 
the calculations of $h$ and $M(\xi)$, while the value of $\alpha$ affects the Green function in Eq. (\ref{26p}) for determining $(z_{0},u_{0})$ in Eq. (\ref{27}). 
Since the coefficient function $M$ corresponding to the map Eq. (\ref{eq:cusp-bed-map}) contains just one cosine component, the matrices $\mathcal{A}_n$, 
$\mathcal{B}_n$ have sparse entries with relatively simple expressions. This ensures a rapid convergence of the path-ordered exponential series. 
We find that it is sufficient to calculate the PE series up to the third iterated integral. All calculations are carried out using the symbolic calculation 
in Mathematica\textsuperscript{\textregistered}  (see also the file "Supplemental Material A").

To validate the perturbation analysis,  we compare the analytic solutions with the results of numerical simulations of the original Boussinesq system 
in Eq. (\ref{eq5new}) using  COMSOL 6.1 software. For the initial conditions, we use the Boussinesq soliton expressions in the flat bed case with the 
depth being  $h$ in the mathematical plane. We center the initial shape in the middle of the interval $I$. 
For each case, we run the numerical simulation twice: one uses the homogeneous boundary conditions and the other uses the periodic boundary 
conditions. This is to ensure that the boundaries of the working interval $I$ are sufficiently away and do not affect the numerical results. 
In general, the end points of the working interval $I$ are placed at $(60\sim80)L_{sol}$ away from the center of the initial soliton shape. 

\subsection{Results}

\begin{figure}
\centering \includegraphics[width=140mm]{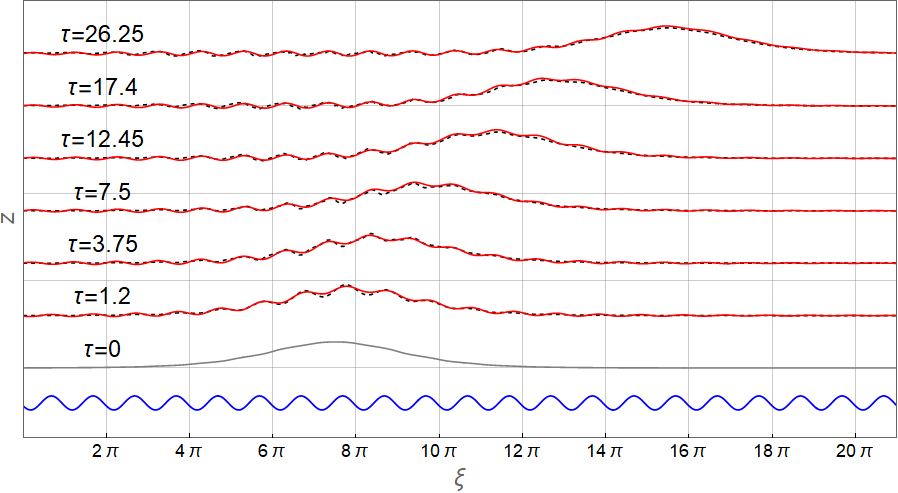}
\caption{Evolution of a perturbed soliton $z_s(\xi,\tau)$ in the mathematical plane, for $\alpha=0.15$, $\beta=0.75$, and $\epsilon = 0.4$. 
{\em Red}: Perturbation solution.  {\em Black dotted:} Numerical simulation. {\em Blue}: Profile shape of $M$ (not at the same vertical scale as 
$z_s$). {\em Gray}: Initial Boussinesq soliton of width $L_{sol}= 7 \pi$ and relative amplitude $\alpha=0.15$.} 
\label{figpoint1upfigpoin2downNEW}
\end{figure}

\begin{figure}
\centering\includegraphics[scale=.5]{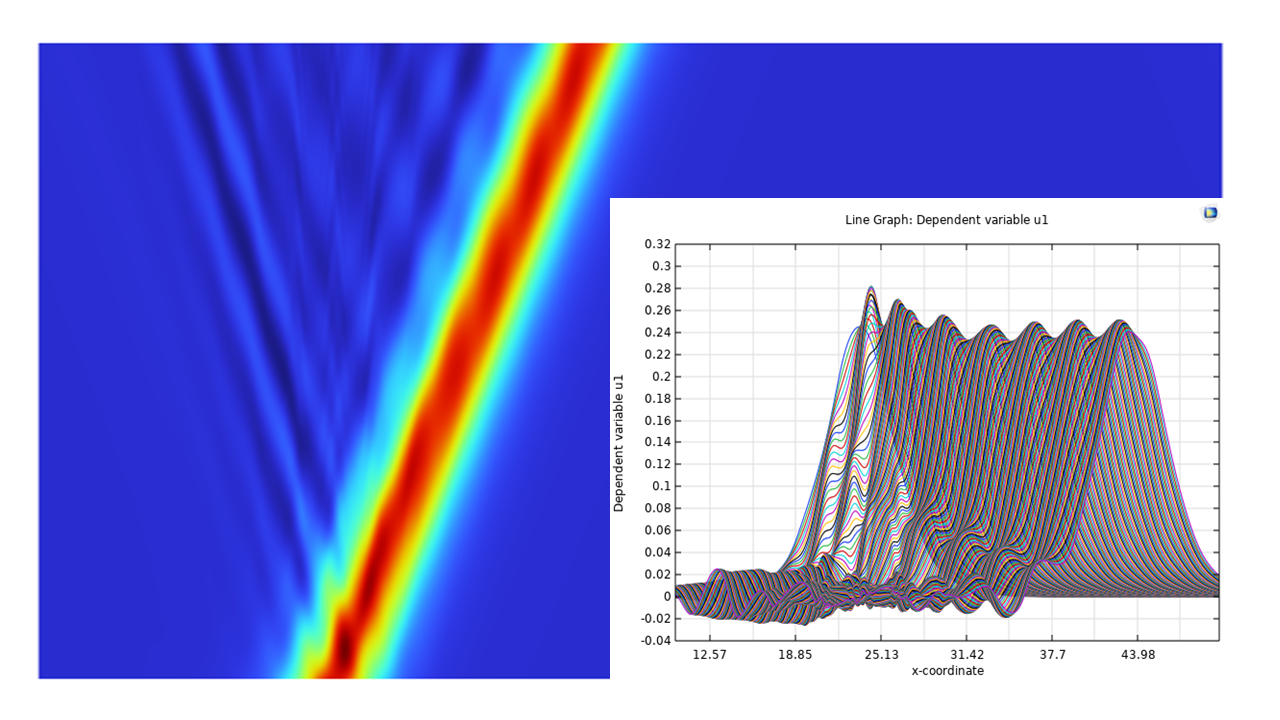}
\caption{Density map showing the soliton evolution and the wake in the soliton tail, from the COMSOL simulation 
for $\alpha=0.25, \beta=0.48$ and $\epsilon=0.263$. The initial soliton has a width $L_{sol}=4\pi$.  Inset: Consecutive phases of the traveling soliton 
in the mathematical plane, showing the decrease and oscillations of the soliton amplitude from numerical calculations.}
\label{figf6}	
\end{figure}

In Figs. \ref{figpoint1upfigpoin2downNEW}-\ref{figf6} we present examples of wave profiles in the mathematical plane, i.e.,  as the functions of $(\xi,\tau)$.
The perturbation solution involves two arbitrary factors: the choice of the matching point between the solution $z_L$ and the perturbed soliton tail, and 
a constant of integration. These  are chosen so that the analytic solution best compare with  the numerical result. 
In Fig. \ref{figpoint1upfigpoin2downNEW}, the large $\beta$ indicates a relatively deep water depth. Thus, the soliton is less influenced by the corrugations, 
and its envelope is less modulated. The high dispersion coefficient in this case also corresponds to an initial soliton of large width and slow velocity.  
The soliton velocity is slightly greater than that of the wave train induced by the corrugations, so the perturbations are lagging behind as time goes on;  
see the relatively smoother soliton front for $\tau > 17.4$ in Fig. \ref{figpoint1upfigpoin2downNEW}. 
The trace of the perturbed soliton, from the COMSOL simulation, is shown in Fig. \ref{figf6} for $\alpha=0.25, \beta=0.48$ and 
$\epsilon=0.263$. As the soliton travels, the wake waves are generated on the soliton tails, while the soliton front is nearly unaffected. 
It is notable that the soliton amplitude decreases due to energy leaking into the wake.  


\begin{figure}
\centering\includegraphics[scale=.43]{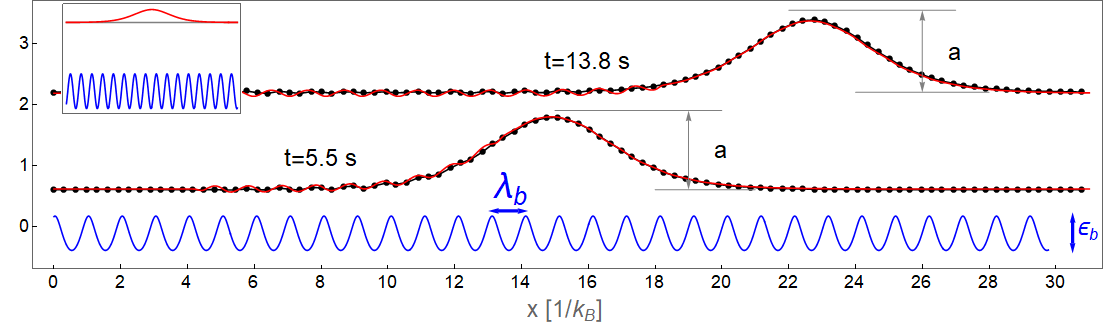}		
\caption{Solution $z_{s}(x, t)$ at two moments of time. Parameters: $\alpha=0.15$,  $\beta=0.6$,  $\epsilon=0.1$, and $L_{sol}=14 \lambda_b$ for 
the initial soliton. {\em Red}: Analytical solution. {\em Black dotted}: Numerical simulation. {\em Blue}: bed corrugations (not at the same vertical scale as 
the waves). \textit{Inset}: The solution shown at the faithful horizontal and vertical scales.}
\label{FIG8}
\end{figure}

\begin{figure}
\centering
\includegraphics[width=120mm]{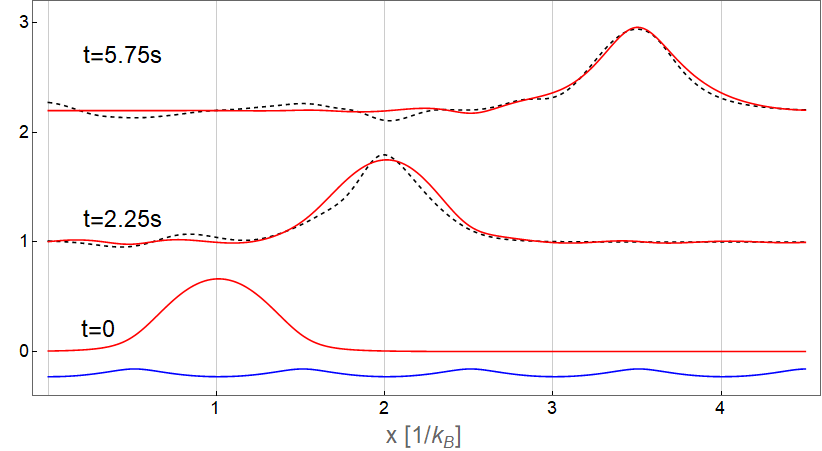}\llap{\parbox[b]{22.3cm}{(a)\\\rule{0ex}{5.9cm}}}\\[10pt]		
\includegraphics[width=120mm]{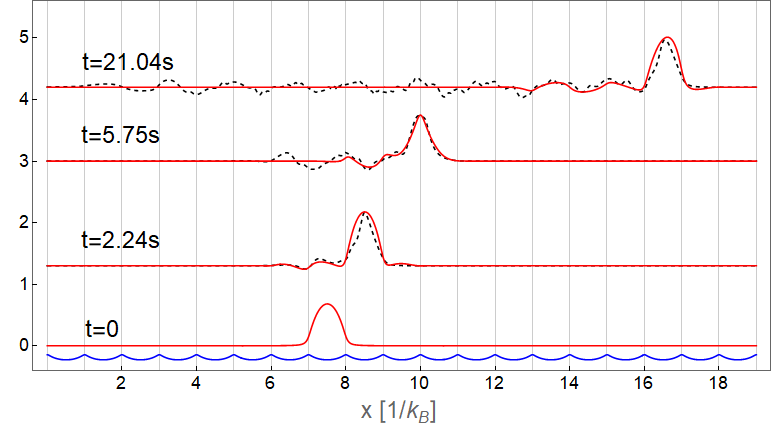}\llap{\parbox[b]{22.3cm}{(b)\\\rule{0ex}{5.9cm}}}		
\caption{Solution $z_s(x,t)$ at various times for $\alpha=0.75$, $\beta=0.35$, and $L_{sol}=3 \lambda_b /2$. 
\textit{Red}: Analytical solution. \textit{Black}: Numerical simulation. \textit{Blue}: Corrugations. 
\textit{(a)} Low corrugation $\epsilon=0.2$. \textit{(b)} High corrugation $\epsilon=0.4$.  }
\label{FIG16}
\end{figure}

\begin{figure}
\centering
\includegraphics[width=120mm]{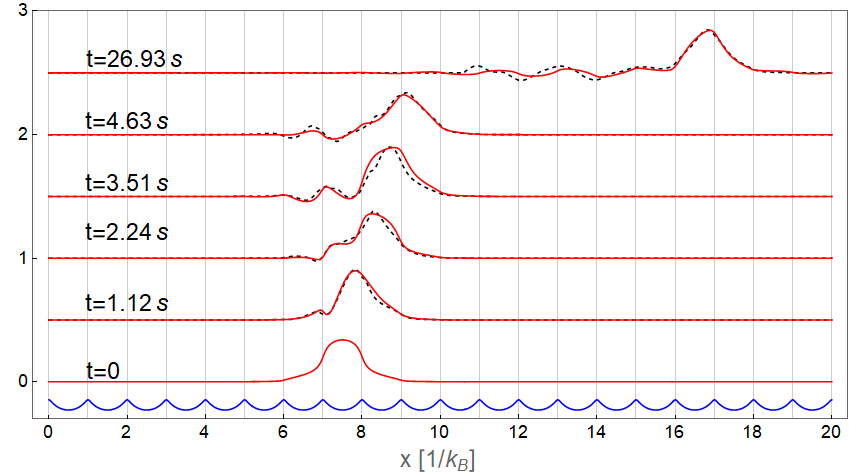}\llap{\parbox[b]{22.3cm}{(a)\\\rule{0ex}{5.9cm}}}\\[10pt]		
\includegraphics[width=120mm]{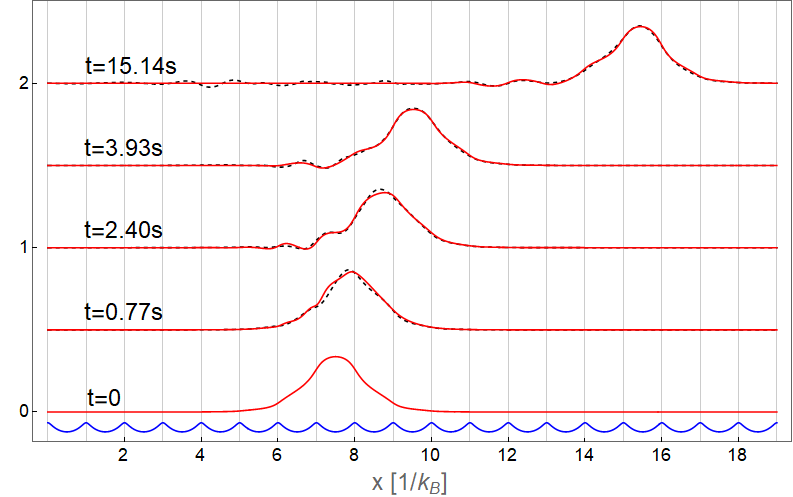}\llap{\parbox[b]{22.3cm}{(b)\\\rule{0ex}{6.8cm}}}		
\caption{Solution $z_s(x,t)$ for $\alpha=0.35$: 
(a) Unstable traveling soliton of  $L_{sol}=3 \lambda_b$ and $\beta=0.35$ over high corrugations of $\epsilon=0.4$. 
(b) Stable traveling soliton of $L_{sol}=5 \lambda_b$ and $\beta=0.75$ over low corrugations of  $\epsilon=0.2$. 
\textit{Red:} Analytic solution. \textit{Black dotted:} Numerical simulation. \textit{Blue}:  Bed corrugations.}
\label{FIG14}
\end{figure}

\begin{figure}	
\centering\includegraphics[width=140mm]{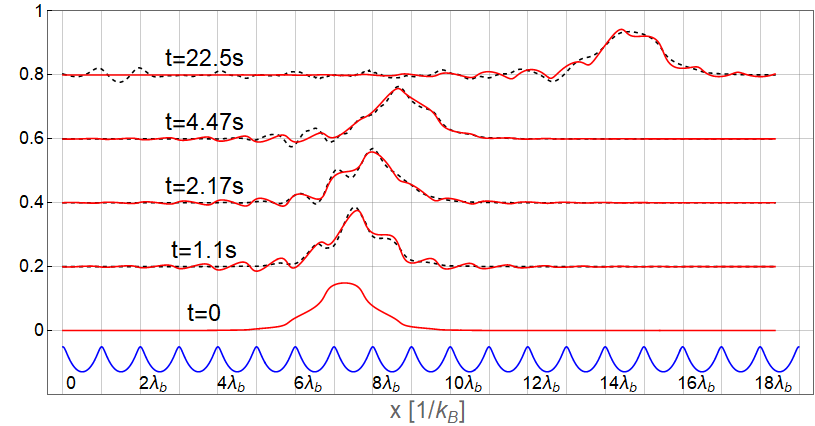}		
\caption{Solution $z_s(x,t)$ for $\alpha=0.15$, $\beta=0.15$, $\epsilon=0.4$.  $L_{sol}=5 \lambda_b$.  
\textit{Red:} Analytic solution. \textit{Black dotted:} Numerical simulation. \textit{Blue}: Bed corrugations.}
\label{FIG9}
\end{figure}

\begin{figure}
\centering
\includegraphics[width=120mm]{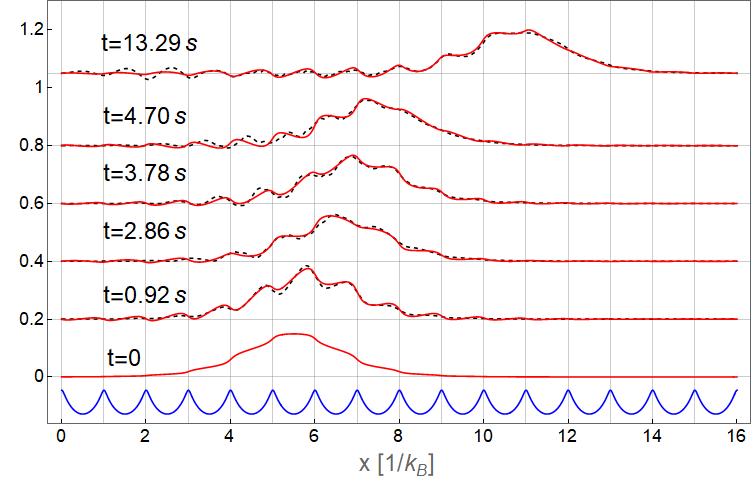}\llap{\parbox[b]{1.5cm}{(a)\\\rule{0ex}{6.9cm}}}\\[10pt]	
\includegraphics[width=120mm]{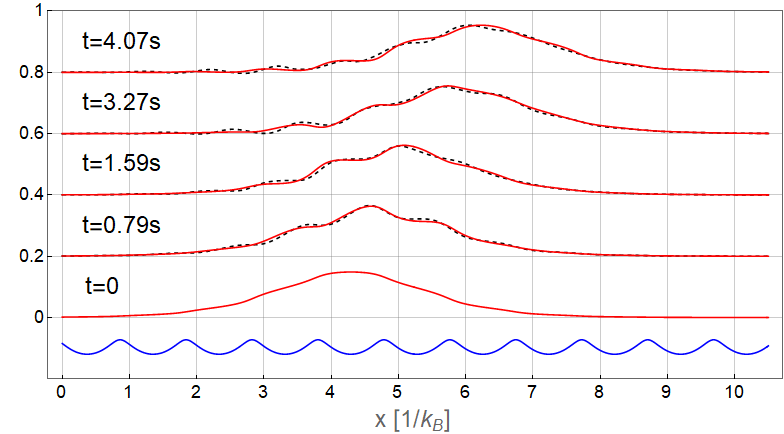}\llap{\parbox[b]{1.5cm}{(b)\\\rule{0ex}{5.8cm}}}		
\caption{$z_s(x,t)$ for  $\alpha=0.15$, $\beta=0.35$, and  $L_{sol}=5 \lambda_b$. (a) $\epsilon=0.4$. (b) $\epsilon=0.2$. 
 \textit{Red:} Analytic solution. \textit{Black dotted}: Numerical solution. \textit{Blue}: Bed corrugations.}
\label{FIG10}
\end{figure}

\begin{figure}
\centering 
\includegraphics[width=105mm]{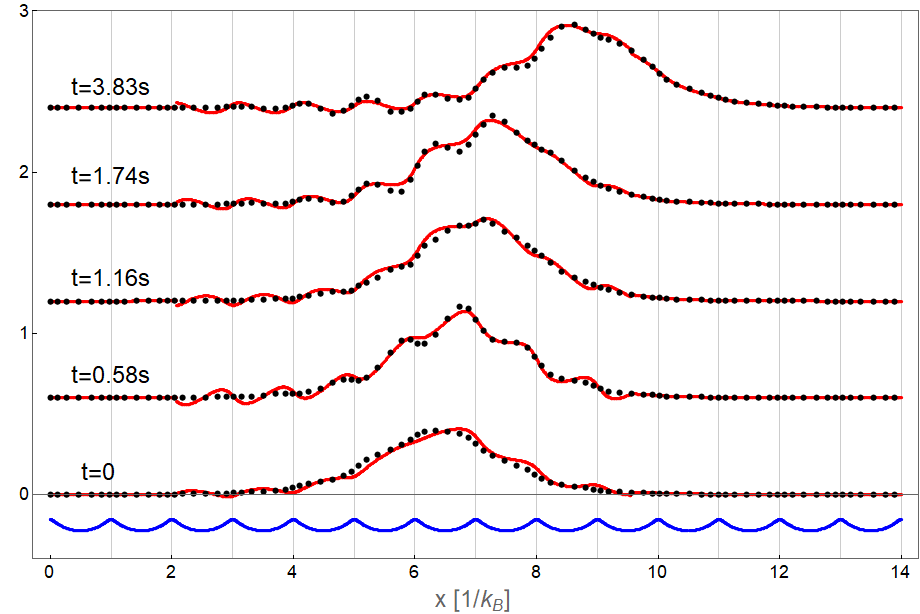}\llap{\parbox[b]{1.5cm}{(a)\\\rule{0ex}{5cm}}}\\[5pt]
\includegraphics[width=105mm]{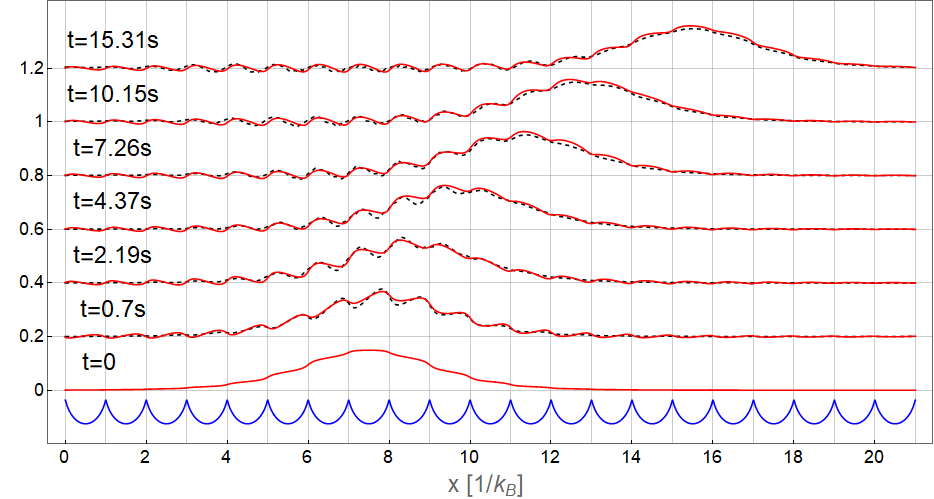}\llap{\parbox[b]{1.5cm}{(b)\\\rule{0ex}{5cm}}}\\[5pt]	
\includegraphics[width=105mm]{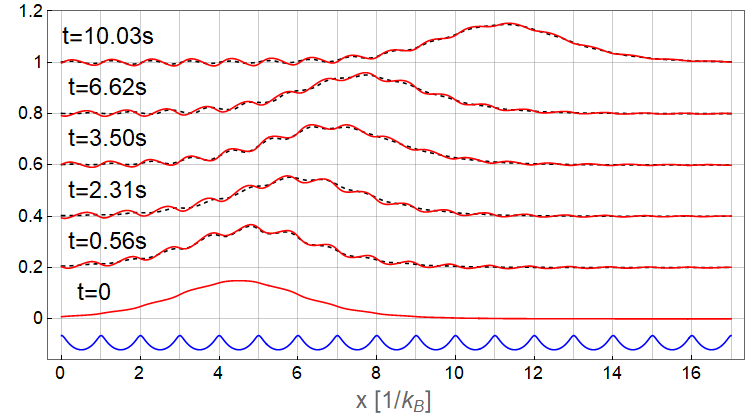}\llap{\parbox[b]{1.5cm}{(c)\\\rule{0ex}{5.1cm}}}\\[5pt]
\includegraphics[width=105mm]{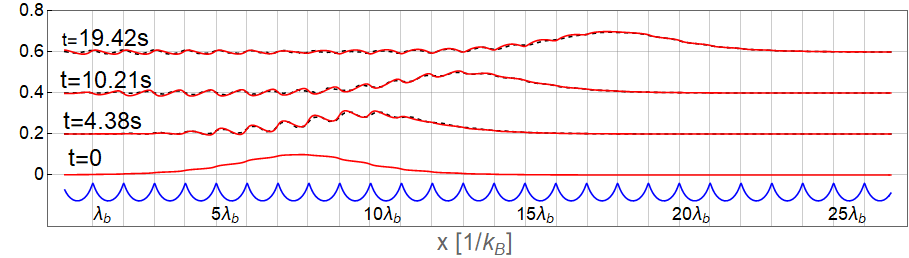}\llap{\parbox[b]{1.5cm}{(d)\\\rule{0ex}{2.35cm}}}
\caption{Stable traveling solitons. 
(a) $\alpha=0.2$, $\beta=0.5$, $\epsilon=0.3$,  $L_{sol}= 5\lambda_b$. 
(b) $\alpha=0.15$, $\beta=0.75$, $\epsilon=0.4$, $L_{sol}=7 \lambda_b$.
(c) $\alpha=0.15$, $\beta=0.75$, $\epsilon=0.2$, $L_{sol}=7 \lambda_b$.
(d) $\alpha=0.2$, $\beta=0.6$, $\epsilon=0.39$,  $L_{sol}=10 \lambda_b$. 
\textit{Red:} Analytic solution. \textit{Black dotted}: Numerical simulation. \textit{Blue}: Bed corrugations.}
\label{fig:stable-solitons}
\end{figure}

\begin{figure}
\centering\includegraphics[width=130mm]{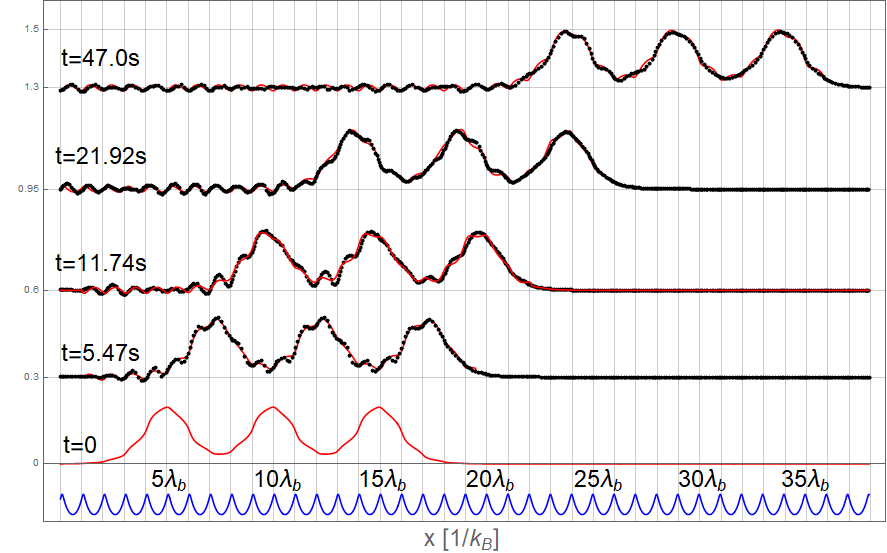}		
\caption{Evolution of a Boussinesq cnoidal wave imitating three solitons for $\alpha=0.2$, $\beta =0.5$ and $\epsilon=0.3$. 
 \textit{Red:} Analytic solution. \textit{Black dotted}: Numerical solution.  \textit{Blue}: Bed corrugations. }
\label{FIG18}
\end{figure}

Hereafter, we present a number of examples in Figs. (\ref{FIG8}-\ref{FIG18}) for various conditions characterized by $\alpha$, $\beta$ 
and $\epsilon$, comparing the perturbation solutions with the results from COMSOL simulations. 
To have a better understanding of the evolution in the physical plane, we transfer $z_s(\xi,\tau)$ to $z_s(x,t)$ using the $\xi$--$x$ relation given 
in Eq. (\ref{eq:x-xi-at-eta0}).  In all figures,  the water depth is not shown, but we plot the shape of bed corrugations (blue curves) blow 
the solutions (red and black curves) for purpose of visualization. The horizontal axis is in the units of $1/k_B$. 
The initial condition is the traditional ``Gauss-bell" shaped soliton in the mathematical plane. However, because of the transformation from $\xi$ to $x$, 
what we call a soliton in the physical space has a quite different shape from the ``Gauss-bell" shape, having multiple local maxima depending on 
the ratio of soliton width to the bed period $\lambda_b$ and corrugation height. When the conditions given by $\alpha,\beta, \epsilon, L_{sol}$ 
concur to produce high stability in time, such localized envelope can travel almost unchanged, exactly like a traditional soliton, 
despite its non-traditional shape.

Fig. \ref{FIG8} shows a very wide ($L_{sol}=14\lambda_b$) soliton of low amplitude traveling over relatively low corrugations. 
The soliton shape is stable in time, though it is 
modulated. The wake is limited because of the rapid attenuation of the perturbation on the soliton tail. The soliton front is free from the 
perturbation. If we take the mean water depth $h_0=0.3$ m, the values of $\alpha$, $\beta$ and $\epsilon$ in this case correspond to the 
corrugation period $\lambda_b=1.22$ m, and initial soliton amplitude $a=0.045$ m. 

Fig. \ref{FIG16} illustrates the cases when the soliton width is narrow,  i.e.,  $L_{sol}= 1.5 \lambda_b$, considering high 
and low corrugation height.  
The high nonlinearity $\alpha=0.75$ suppresses the effect of perturbation on the tall soliton, as long as the corrugations are not so high (e.g., $\epsilon<0.2$). 
The analytic solutions of the trailing waves compare reasonably well with the numerical results up to $t=3.4$ s. As later times, the analytical solution 
gives a faster decaying trailing waves, compared to the numerical solution. While the soliton height compares reasonably well, there are times when 
the soliton given by the numerical solution has a  narrower shape than that given by the analytical solution; see $t=2.24, 2.25$ s. 
Since the linear water waves travel at $5.4$ m/s which is less than the soliton velocity of $6.78$ m/s,  the trailing waves do not catch up with the soliton. 
This is clearly indicated by the nearly perturbation-free soliton front. For the case of higher corrugations in Fig. \ref{FIG16}(b),  
stronger deformation of the soliton tail can be seen for $t\ge 5.75$ s. Nevertheless, the strong nonlinearity $\alpha$ restores the stability of the soliton, 
and its shape is reconstructed at $t=21.04$ s. 

These results align with the calculations and findings in \cite{2009Craig}, where wave propagation over a randomly varying bottom also leads to scattered wave fields in addition to the primary KdV soliton component. As shown in Figs. 5-6 of \cite{2009Craig}, the main coherent solutions persist, retaining their fundamental properties of momentum and energy transport, much like in the classical flat-bottom problem. In our analysis, we observe, consistent with the conclusions in \cite{2009Craig}, that the scattering is not substantial enough to cause significant dispersion of the principal coherent solutions, within the approximation regime used in these results.

The strong scattering effects behind the soliton in Figs. \ref{FIG16} exhibit similar features with a  Bragg resonance \cite{2009Craig,2007Guy,Jie2012,Jie2007} since the soliton width and corrugation wavelength are comparable in this case. However, in the case of a soliton, the surface wavelength is infinite and it is not clear if the soliton width can be used in place of the surface wavelength in the $m\lambda = 2\lambda_b$,  $m = 1,2,\dots$ Bragg condition \cite{Jie2012,Jie2007}. Consequently, we do not think that the scattered waves in this case are due to the Bragg scattering, since strong perturbation waves are similarly observed for wider solitons as one can notice in the following Figs.  \ref{FIG14}-\ref{fig:stable-solitons}. On the other hand, since the water continues to exhibit weaker (shallower amplitude) nonlinear oscillations and waves after the soliton perturbation has subsided, as shown in section \ref{subsec4}, this scattered waves could be related to a local Bragg resonance effect.  We will consider the study of this specific effect with our perturbation solution Eq. (\ref{linaa}) in a forthcoming paper.

The competing effects of dispersion and bed corrugations are shown in Fig. \ref{FIG14}, considering two situations: (a) a narrow, mildly dispersive soliton 
over high corrugations, and (b) a wider, more dispersive soliton over lower corrugations. In both situations, the nonlinearity $\alpha =0.35$. 
In Fig. \ref{FIG14}(a), the soliton shape is highly modulated by the perturbation and does not recover to its original form even after a long time. This is 
an unstable traveling soliton. 
In Fig. \ref{FIG14}(b) the smaller corrugations generate notable amount of modulation of the soliton envelope at $t=2.40$ s, but the soliton shape approximately recovers at $t=15.14$ s due to the increased dispersion. The traveling soliton is stable similar to the cases studied in \cite{2009Craig}.  Recall that $\beta=(k_B h_0)^2$. Thus, a larger $\beta$ can be obtained by either increasing the mean water depth $h_0$ or shortening the corrugation 
period $\lambda_b=\pi/k_B$, or both.  
For instance,  for $h_0=0.3$ m, $\beta =0.35$ in Fig. \ref{FIG14}(a)  corresponds to $\lambda_b=1.59$ m, and $\beta = 0.75$ in Fig. \ref{FIG14}(b) 
is for  $\lambda_b=1.08$ m. The results in Fig. \ref{FIG14} indicate that wider solitons in general have their envelopes less perturbed. A wider soliton 
essentially `feels' the reduction of water depth due to the relatively dense corrugations, but does not `feel' the individual corrugations. 
In both situations,  the analytic solutions compare reasonably well with the numerical results for both the soliton envelope and the trailing waves. 
For very weak dispersion $\beta = 0.15$ but relative high corrugations of $\epsilon=0.4$, the modulation of the soliton envelope is so strong that 
multiple local maxima appear; see $t\ge1.1$ s in Fig. \ref{FIG9}. Furthermore, the soliton front is affected. This effect is in agreement with the results presented  in \cite{2007Guy} (for example in Fig. 4.3) where the soliton wave becomes sharper when shoaling. The soliton tends to be unstable for  $t > 5$ s. 
One reason for the strong deformation is that the velocity of linear waves is $3.43$ m/s and exceeds the soliton velocity of  $3.2$ m/s. 
Thus, the perturbation waves  take over the soliton.  For this configuration of parameters, the amplitude of the wave slowly decreases with time.
To better illustrate the effect of bathymetry, we show the soliton modulations by varying $\epsilon$ while keeping $\alpha$, $\beta$, and $L_{sol}$ 
unchanged; see Fig. \ref{FIG10}. As expected, higher corrugations induce greater modulation of the soliton and longer train of waves in its wake. 
The amplitude of the wave decreases in time. A remarkably good agreement is seen between the analytic and numerical solutions.  

Fig. \ref{fig:stable-solitons} shows  that the soliton evolution exhibits a very good stability in time for a wide range of parameters. In those cases, 
the amplitude of the initial soliton is small with $\alpha=0.15\sim0.2$. The soliton width relative to the bed period is increased from 
$L_{sol}/\lambda_b = 5$ to 7 and 10. In all cases, the solitons and the 
trailing waves are both stable. This demonstrates that the soliton width is not necessarily a strong criterion for stability.  
In Fig. \ref{FIG18}  an example is given using an initial condition of three-soliton cnoidal waves, imitating three solitons for a condition of 
weak nonlinearity, medium dispersion and bed corrugations. The waveform propagates for a long time ($47$ s), nearly unchanged.  
In all these stable cases, the analytical and numerical solutions are in very good agreement, not only for the soliton shape but also 
the trailing  waves. 

We note that the height of the perturbed soliton slight varies as the soliton propagates; see, e.g., the inset of Fig.\ref{figf6}. This can be explained by the conservation of total area under the perturbed solution, resulting from the Hamiltonian Eq. (\ref{eq30old}). In some cases, from the solutions at longer times 
we observe a very weak, very low frequency modulation of the (otherwise monochromatic) trailing waves in the wake. 
This behavior may be caused by a Doppler effect. Waves away from the soliton support stabilize at a certain phase velocity, while the waves close to the 
trailing front of the soliton reflect at the moving boundary. 
As we have shown in Section \ref{subsec4}, since the two velocities $v_{ph}$ and $v_{sol}$ have close values,  the interference of waves may result in low frequency modulation beats.  As a summary, we note that the analytic solutions compare well with the numerical solutions for a large range of values for 
$\alpha,\beta, \epsilon$. The agreement is less satisfactory when the soliton width becomes shorter than the period of  bathymetry.   

The analytic solutions for the perturbations $z_N, u_N$  allow for some qualitative analysis of these nonlinear waves. For the large majority of cases, all the terms in Eqs. (\ref{eb5pp}, \ref{eb5ppp}) are iterative integrals of positive functions given by the soliton shape or velocity profile, or both   Eqs. (\ref{soliton}). Such consecutive integrations render strictly monotonic functions at every step $k$ of integration. Because the lower limits of these integrals are chosen at the soliton front (causality condition for retarded waves propagation) it results that the main contribution of these perturbation integrals is localized in the trailing part of the soliton. This effect is visible in the large majority of the figures. There are a few exceptions noticed when the soliton half-width is around $L_{sol}\simeq 5 \lambda_{B}$. In Figs. \ref{FIG9}, \ref{FIG10}(a), and \ref{fig:stable-solitons}(a) the soliton envelope becomes modulated even at its leading front. A similar effect of weaker intensity is also noticeable in  Figs. \ref{FIG14}(b), \ref{FIG10}(b), and Figs. \ref{fig:stable-solitons}(b) and (c), because in these cases either corrugation is smaller ($\epsilon=0.2$) or $L_{sol}=7 \lambda_B >5 \lambda_B$. At this critical value $5 \lambda_{B}$ for the half-width the convergence of the series of iterated integrals in Eqs. (\ref{eb5pp}, \ref{eb5ppp}) is weaker, which could be a signature of a nonlinear Bragg resonance. Indeed, from Eq. (\ref{eqvv}) our model for scattered waves Eq. (\ref{eqv}) predicts a reverse of direction in the group velocity when $L_{sol}\simeq 5 \lambda_B, \alpha\simeq 0.18, \beta\simeq 0.4$, situation which is very close to the quasi-resonant cases mentioned above.

\subsection{Stability}

Because the coefficients of the time derivative and triple $\xi$ derivative have the same sign in Eq. (\ref{eqp1}) and Eq. (\ref{eq5new}a), these systems belong to the class of ill-posed differential equations \cite{illposed}. In general, classical solutions to these equations with arbitrary initial data are not expected to exist for large intervals of time except for some special choices of the initial data. For some of the solutions, a high growth rate of waves with short wavelengths may occur. At a long time scale these short-waves become dominant in the solution due to the nonlinearity of the equation and numerical instability. 
Therefore, we do not expect the approximate analytic models developed in Section \ref{sec3} and \ref{sectionStefanKdV} to accurately represent  the 
evolution of waves in  long time intervals. Here we present a stability analysis for the evolution of perturbed soliton amplitude, based on the numerical simulations. 

\begin{figure}
\centering\includegraphics[scale=.5]{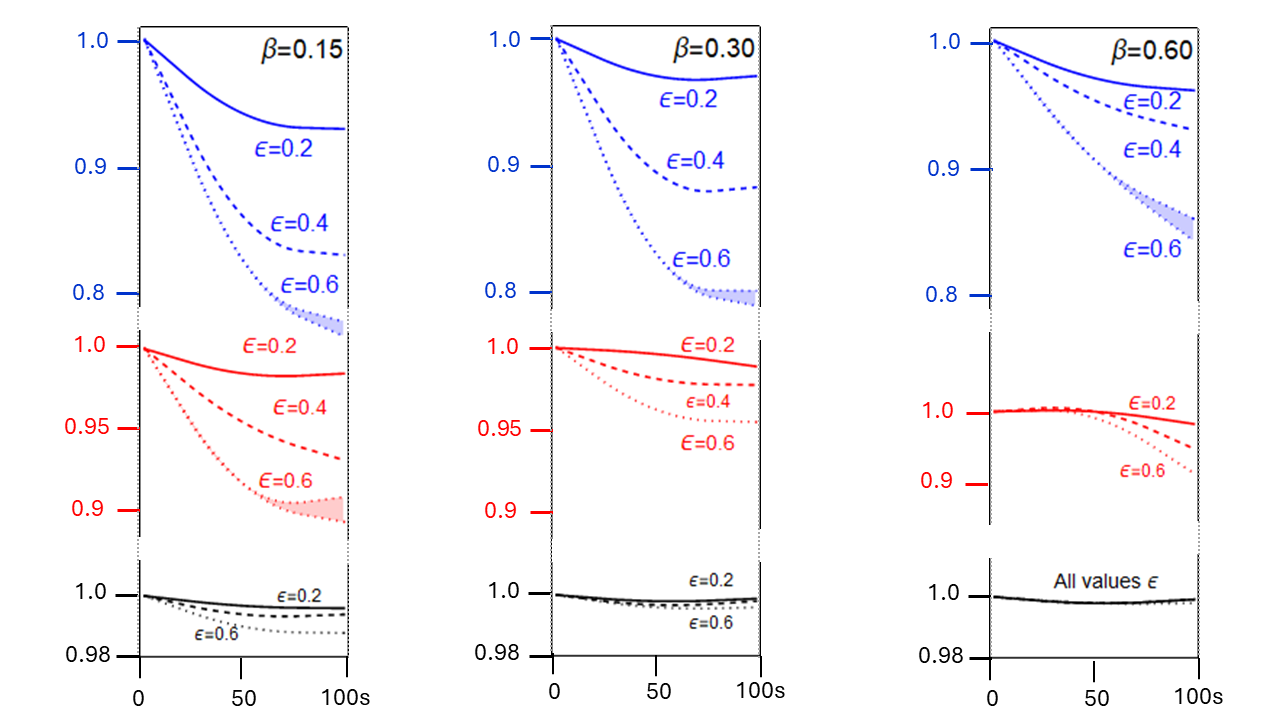}\llap{\parbox[b]{16.5cm}{(a)\hspace{5.0cm}(b)\hspace{5.5cm}(c)\\\rule{0ex}{9.25cm}}}
\caption{Evolution of soliton amplitude in time (seconds) for various parameters, based on numerical simulations. 
The amplitude of the perturbed soliton is normalized by the initial amplitude (i.e., $\alpha=a/h_0$). 
For the dispersion parameter: (a) $\beta=0.15$, (b) $\beta=0.30$, (c) $\beta=0.60$. 
For the nonlinearity (or initial soliton height): black, $\alpha = 0.15$; red, $\alpha=0.3$; blue, $\alpha=0.6$.  
The shaded areas between curves of the same color indicate moments of soliton fragmenting by instability. 
For the corrugation height: solid curve, $\epsilon = 0.2$; dashed curve, $\epsilon=0.4$; dotted curve, $\epsilon=0.6$. }
\label{figparam2}
\end{figure}
In Fig. \ref{figparam2}, the long-time evolution of the amplitude of the perturbed solitons is shown for various conditions. 
The general behavior of solitons shows decreasing amplitude due to dispersing their energy into the perturbation waves in the soliton wake.  
Obviously, high corrugations cause more rapid attenuation of soliton amplitude. 
The decrease in amplitude and strong modulation of the soliton envelope at earlier times are clearly seen in  Figs. \ref{FIG9}--\ref{FIG10} for 
$\epsilon=0.4$. 
For small solitons (black curves for $\alpha = 0.15$ in Fig. \ref{figparam2}), the temporal variation of the normalized soliton amplitude is not 
very sensitive to the corrugation height $\epsilon$ unless the dispersion is weak (e.g., $\beta =0.15$). 
For taller solitons with stronger nonlinearity, the attenuation rate of normalized soliton amplitude increases rapidly with $\epsilon$. 
The effect of dispersion tends to compensate the effect of bed corrugations, since soliton energy attenuation becomes weaker as $\beta$ increases. 
The shaded areas between the curves of the same color indicate the moments when the soliton is fragmenting by the instability. 
Numerical simulations for various cases indicate that the velocity  $u_{sol}$ and  width $L_{sol}$ of the soliton are less affected, changing no more 
than  $2-3 \%$ within the first $50\sim80$ seconds of traveling. 
It is possible that some of the results may be slightly affected by numerical procedures, especially by the imposed periodic boundary 
conditions, even though the boundaries are always placed at distances about $(50\sim80) L_{sol}$ away from the center of the initial pulse. 
These results, including  the stability of soliton shapes and the effects of corrugations on soliton amplitude and velocity,  are 
consistent with the studies of similar KdV soliton waves traveling over various types of bathymetries in \cite{101,104,nako}.

\section{Boussinesq system  as a perturbed KdV equation}
\label{sectionStefanKdV}

Setting  $Mu_{s}=w_{\xi}$ and substituting it into Eq. (5b), we obtain
\begin{equation}\label{2}
z_{s}=-w_{\tau}-\frac{\alpha}{2M^2}w_{\xi}^2.
\end{equation}
Inserting into Eq. (5a)  we have 
\begin{equation}\label{3}
Mw_{\tau \tau}-hw_{\xi \xi}-\frac{\beta h^3}{3}w_{\xi \xi \xi \xi}+\frac{\alpha}{M}w_{\xi }w_{\xi \tau}+\alpha\left(\frac{w_{\tau}w_{\xi}}{M}\right)_{\xi}+\frac{\alpha^2}{2}\left(\frac{w_{\xi}^3}{M^3}\right)_{\xi}=0.
\end{equation}
Using $M(\xi)$ in  Eq. (\ref{eqM}) and for $\epsilon<1$,  we re-write Eq. (\ref{3}) into 
\begin{eqnarray}
&& w_{\tau \tau}-hw_{\xi \xi}-\frac{\beta h^3}{3}w_{\xi \xi \xi \xi}+\alpha w_{\xi} w_{\xi \tau}+\alpha\left( w_{\tau}w_{\xi} \right)_{\xi} +\frac{\alpha^2}{2}\left(w_{\xi}^3\right)_{\xi}\nonumber\\
&& \qquad +\epsilon \big( -M_N w_{\tau \tau}+2\alpha M_N w_{\xi} w_{\xi \tau}+\alpha M_{N,\xi} w_{\tau} w_{\xi} +\alpha M_N w_{\tau} w_{\xi \xi}\big) \nonumber\\ 
&& \qquad +\epsilon\Big(\frac{3\alpha^2}{2} M_{N,\xi} w_{\xi}^3 +\frac{9\alpha^2}{2}M_N w_{\xi}^2 w_{\xi \xi} \Big)=0.
\label{7}
\end{eqnarray}
Let us consider the approximation of long waves in the slow variables
\begin{equation}\label{8}
X=\epsilon^{1/3}(\xi+\sqrt{h}\tau), \ T=\frac{\beta \epsilon h^{5/2}}{6} \tau,
\end{equation}
and consider $w(\xi, \tau)\rightarrow w(X,T)$ as the field of  nonlinear interaction  balancing the dispersion effect. We then have from Eq. (\ref{7})
\begin{equation}\label{10}
\frac{\beta h^3}{3}\biggl( w_{XT}+w_{XXXX}+\frac{9\alpha}{\beta h^{5/2}}w_{X}w_{XX}\biggr) =\epsilon^{7/3} M_N w_{XX}.
\end{equation}
Denoting  $w_{X}(X,T)=\Psi(X,T)$, we have essentially a perturbed KdV equation  in the form
\begin{equation}\label{11}
\Psi_{T}+\Psi_{XXX}+\gamma \Psi \Psi_{X}=\delta m_1(X) \Psi_{X},
\end{equation}
where $\delta$ is an arbitrary scaling for the moment, and 
\begin{equation}\label{12}
\gamma=\frac{9 \alpha}{\beta h^{5/2}}, \ \  m_1=\frac{3}{h^2 \beta}M_N.
\end{equation}
We can apply a classical perturbation calculus to Eq. (\ref{12}). If $\gamma=0$ and $\delta^2=\epsilon$ we are in the situation as  in  \cite{Jie2019revisit}. Following the assumptions made in that work  we then find the linear perturbed equation
$\Phi_{XT}+\Phi_{XXXX}+M_N h \Phi_{XX}=0$. Back to the KdV case in Eq. (\ref{11}) and using the same notations as in \cite{malom}, we have 
\begin{equation}\label{14}
U_{T}-6 UU_{X}+U_{XXX}=\epsilon^{\frac{1}{3}} m_1 (X) U_{X},
\end{equation}
where we use the transformation
\begin{equation}\label{15}
U(X,T)=-\frac{3 \alpha \Psi(X,T)}{2 \beta h^{5/2}},
\end{equation}
following the general perturbation theory that refers to equations in the form of $U_{\tau}-6 U U_{\xi}+U_{\xi \xi \xi}=\delta P[U]$ as the perturbed KdV equation. The soliton of the unperturbed KdV equation has the form
\begin{equation}\label{17}
U(X,T)=-2 k^2 \hbox{sech}^2 \theta, \ \ \theta=k(X-\zeta), \ \zeta=4 k^2 T+\zeta_0, 
\end{equation}
with the arbitrary parameter $k$ labeling the soliton family. The perturbation modifies the soliton parameters in the following way \cite{malom} 
\begin{equation}\label{18}
\frac{dk}{dT}=-\frac{\delta}{4k}\int_{-\infty}^{\infty}P[U] \hbox{sech}^2 \theta d\theta,
\end{equation}
\begin{equation}\label{19}
	\frac{d \zeta}{d\tau}=4k^2-\frac{\delta}{4 k^3}\int_{-\infty}^{\infty} P[U]\biggl( \theta +\frac{1}{2}\sinh 2 \theta \biggr) \hbox{sech}^2 \theta d \theta .
\end{equation}
In our case the perturbation $P[U]=m_1 (X) U_{X}$ has a periodic factor, which is expressed in a Fourier series
\begin{equation}\label{20}
m_1=\frac{3}{h^2 \beta}\biggl[
-\frac{2 \pi h}{\lambda_B} 
\biggl( 
	\sum_{j=1}^{N_c}
	\frac{j b_j}{\hbox{sinh}  \frac{2 \pi j h}{\lambda_B}} \cos (2 j X) +\frac{j   c_j}{\hbox{sinh} \frac{2 \pi j h}{\lambda_B}} \sin (2 j X)
	\biggr) 
	\biggr].	
\end{equation}
To simplify the computations, we use the following notations
\begin{equation}\label{21}
	B_j=\frac{j b_j}{\hbox{sinh}  \frac{2 \pi j h}{\lambda_B}}, \ \ C_j=\frac{j c_j}{\hbox{sinh}  \frac{2 \pi j h}{\lambda_B}}.
\end{equation}
Substituting  Eq. (\ref{20}) into the integrals in Eqs. (\ref{18}, \ref{19}), we obtain 
\begin{equation}\label{22}
\frac{dk}{dT}=\delta \sum_{j=1}^{N_c} \frac{4 C_j j^2 (j^2+k^2) \pi^2 \ \hbox{cosech} \frac{j \pi}{k}}{h k^2 \beta \lambda_B}
\end{equation}
and
\begin{equation}\label{23}
\frac{d \zeta}{dT}=4 k^2-\frac{2 \pi^2 \delta}{h k^4 \beta \lambda_B} \sum_{j=1}^{N_c} B_j j  \biggl[-4 k^3+j (j^2+k^2)\pi \hbox{coth}\frac{j \pi}{k} \biggr] \ \hbox{cosech}\frac{j \pi}{k}.
\end{equation}
Eq. (\ref{22}) is a highly nonlinear differential equation for the scaling factor $k(T)$. For example, retaining in the sum the terms up to  $j=4$ and integrating, we obtain a modulation of the amplitude of the traveling soliton similar to the result from numerical integration of  Eqs. (\ref{eq5new}) as shown in the  inset of Fig. \ref{figf6}. An interesting observation is that for the periodic bed that leads to $ M_N(\xi)\sim \cos 2 \xi$,  the derivative $ dk/dT$ cancels and correspondingly the amplitude and  velocity of the perturbed soliton do not change at the first order. Only the soliton phase is changed by the perturbation. This  is in agreement with  the approximate analytic solutions and the numeric results presented in Section III and IV.

\section{Conclusions}
\label{secconcl}

We have investigated the perturbations induced by a periodic bathymetry on traveling solitons in a two-dimensional configuration, using the 
terrain-following Boussinesq model for weakly nonlinear and weakly dispersive waves developed in \cite{Jie2019revisit}. 
The model applies to periodic bed bathymetry of arbitrary profile shape and height, provided the bed is fully submerged. 
Assuming that the height of bed undulations is small compared to the mean water depth, characterized by the relative height $\epsilon$, we have 
developed a perturbation solution to the nonlinear Boussinesq model. The leading order term of the solution is the classical flat-bed traveling Boussinesq soliton. 
The $\mathcal{O}(\epsilon)$ term of the solution describes the modulation of the soliton envelope. The equations for the $\mathcal{O}(\epsilon)$ perturbations 
are of Hamilton type, linear and non-homogeneous, in which the influence of bed bathymetry is transferred to the non-homogeneous terms.  
The solution in the form of modulated traveling waves is constructed using the Fourier series, and solved using the Green function method and 
path-ordered exponential method. We have proven the convergence of the series. In section III.B we shown that the scattering  perturbations  are free waves generated 
by the soliton motion in its wake. They are described by the fourth-order dispersive linear wave equations which are homogeneous and can be solved 
directly following the approach of linear Floquet theory \cite{Jie2019revisit, Jie2019waveforms}. The perturbation model also suggests the presence of a nonlinear resonance effect at $L_{sol}\simeq 5 \lambda_B$, consistent with the Bragg wave-bottom resonances observed in reference \cite{2007Guy}.

We have presented a number of examples of the analytical solutions, and for validation, compared them with the results of numerical simulations of the original 
terrain-following Boussinesq system. The examples cover a large range of values of the model parameters, illustrating the effects due to the interactions 
among wave nonlinearity, dispersion, and bed variations. For the large majority of values of the parameters, the analytic solutions  compare 
well with the results of numerical simulations, showing good agreements both in the perturbed soliton envelope and the secondary trailing waves induced by 
the soliton motion in its wake. 

We have also given a comprehensive examination of the stability of the perturbed traveling solitons. 
At last, we presented a procedure to map the Boussinesq system under study into a perturbed KdV equation, and obtained its corresponding 
perturbed soliton. We compared this solution with the above-mentioned perturbation solution, noting a very good agreement. 

The analytic model also offers advantages for the qualitative analysis of the soliton perturbations. Using the perturbation model developed in section III one can approach any type of multi-soliton solutions, including the case of perturbing the collisions of two-solitons, or even piece-wise continuous initial conditions.  Even if numerical approaches for nonlinear evolution equations play a crucial role, the most difficult (and important) problem in the numerical approach is to find a proper discretization procedure which produces a solution that mimics the "physical behavior"  as much as possible. This difficulty increases in the case of nonlinear non-autonomous equations, like in our case. As a similar  example we mention the Riccati equation $y'=y^2+c$ which is analytically solvable, yet numerically implemented generates  a completely chaotic/non-integrable logistic-type dynamics for any finite discretization step. Accordingly it is desirable to rely in parallel on some analytical approaches, including the use of analytical solution with numerically computed coefficients. 

The theoretical approaches presented here are valid for any type of periodic bathymetries, and the method can readily be extended to non-periodic ones.  
Such analytical solutions, though tedious in calculating them, offer the possibility to validate various numerical 
approaches in the literature for similar problems.  
In addition, our approaches pose a methodical value for the general field of theories on nonlinear non-autonomous systems.

\section*{Acknowledgement}
The author A. Ludu would like to thank the US Office of Naval Research (Program ONR-SFRP 2021-2023, Naval Research Laboratory at Stennis Space Center) and to Richard Allard for the support of this project. 

\section{Appendix. Boussinesq solitons over a flat bed}
\label{sec8}
\setcounter{equation}{0}
\renewcommand\theequation{A\arabic{equation}}

\begin{figure}
\centering\includegraphics[scale=.8]{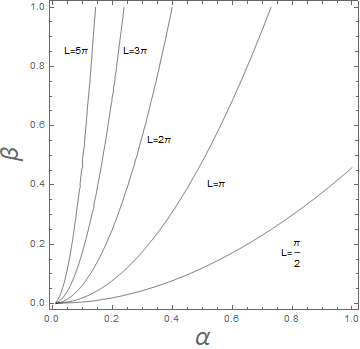}
\caption{Half-width $L_{sol}$ as a function of $(\alpha, \beta)$.}
\label{figsol}
\end{figure}

We describe the procedure to obtain one- or multi-soliton solutions for a flat bottom case, which are used to construct the perturbation solutions 
in Section \ref{sec3}. By applying a B\"{a}cklund transformation
\begin{equation}
\xi \rightarrow X=\frac{\xi}{2h}\sqrt{\frac{3}{\beta}}, \ \ \ v(X,\tau)=-\frac{\alpha u_s}{\sqrt{h}}, 
 \ \ w(X,\tau)=1+\frac{\alpha z_s}{h}-\frac{2 \alpha \sqrt{h \beta}}{\sqrt{3}}u_{s,\xi},
\nonumber
\end{equation}
Eqs. (\ref{eq5new}) with $M=1$ (for a flat bed $\epsilon =0$) become a BK IST-integrable system \cite{1975Kaup,2000Li}
\begin{equation}\label{eq20}
v_{\tau}=\frac{1}{2}(2 w-v_{X}+v^2)_{X}, \ \ w_{\tau}=\Big( v w +\frac{1}{2}w_{X} \Big)_{X}. 
\end{equation}
This is a consequence of the compatibility condition (i.e., the zero-curvature equation) for the associated linear spectral problem. 
To construct explicit soliton solutions, a Miura transform is performed \cite{2003zhang}
\begin{equation}\label{eq25}
q=e^{\int u_s d\xi}, \ \ r=-\Big( 1+z_s -\frac{1}{2}u_{s,\xi}\Big) e^{-\int u_s d\xi},
\end{equation} 
which maps Eqs. (\ref{eq20}) into the first member of AKNS hierarchy system.  A right-moving one-soliton solution to Eqs. (\ref{eq5new}) 
(with $M=1$ for $\epsilon=0$) can then be obtained \cite{2003zhang}, as follows. 
\begin{eqnarray}
z_{sol} &\!=\!&\frac{
	\alpha (4+\alpha^2) \Big[ 2+(2+\alpha^2) \ \hbox{cosh} \ \Big( 
	\frac{\alpha \sqrt{3(4+\alpha^2)}(2 \xi-\tau(2+\alpha^2))}{4\sqrt{\beta}} 
	\Big) \Big]
}{
	\Big[ 2+\alpha^2+2 \ \hbox{cosh} \   \Big( 
	\frac{\alpha \sqrt{3(4+\alpha^2)}(2 \xi-\tau (2+\alpha^2))}{4\sqrt{\beta}} 
	\Big)  \Big]^2
}
\nonumber \\
&\!=\!&\frac{4\alpha}{ 1+ \ \hbox{cosh} \  \frac{\alpha \sqrt{3}(\xi-\tau)}{\sqrt{\beta}}}+\mathit{O}(\alpha^2)
\nonumber \\[5pt]
 u_{sol} &\!=\!&\frac{
	\alpha (4+\alpha^2) }{
	2+\alpha^2+2 \ \hbox{cosh} \   \Big( 
	\frac{\alpha \sqrt{3(4+\alpha^2)}(2 \xi-\tau(2+\alpha^2))}{4\sqrt{\beta}} 
	\Big) 
}
\label{soliton}
\end{eqnarray}
with $u_{sol},  \ z_{sol} \le \alpha$. The traveling velocity $v_{sol}$ of this one-soliton and its half-width $L_{sol}$ are given as 
\begin{equation}\label{eqv}
v_{sol}=1+\frac{\alpha^2}{2}, \ \ L_{sol}=\frac{2 \sqrt{\beta}}{\alpha \sqrt{3(4+\alpha^2)}},
\end{equation}
Both the nonlinearity $\alpha$ and dispersion $\beta$ affect the half-width $L_{sol}$ (Fig. \ref{figsol}). 
It is important to note this dependence since a higher $\alpha$ and smaller $\beta$ may keep the soliton width constant, hence 
having a strong impact on the flow conditions.  
Similar to large majority of solitons, the smaller the soliton amplitude is, the larger the half-width is, and the slower the soliton moves. 
As the nonlinearity $\alpha$ becomes stronger, the soliton becomes narrower. 
This one-soliton has a single peak when its amplitude is less than $2/\alpha$ and double-peak when the wave amplitude is larger than $2 /\alpha$, 
thus having some remarkable features. Since the Boussinesq model is valid for water waves of small amplitude (which is scaled to be $\alpha$ here), 
the double-peaked soliton is not physically meaningful for our situations.  
By applying other types of Darboux transforms, one can also obtain other multi-soliton solutions.

\end{document}